\documentclass[aps,prd,10pt,showpacs,amsmath,showkeys,twocolumn,floatfix,amssymb, preprintnumbers, nofootinbib, superscriptaddress]{revtex4-1} 
\usepackage{epsfig,dcolumn}
\usepackage{graphicx}
\usepackage{comment} 
\usepackage[usenames]{color}
\usepackage{graphicx}
\usepackage{bm}
 \usepackage{ifpdf}
 \usepackage{subcaption}

\usepackage[normalem]{ulem}
\usepackage[dvipsnames]{xcolor}
\usepackage[utf8]{inputenc}
\usepackage{hyperref}
\hypersetup{ 
  pdfnewwindow=true,      % links in new window
  colorlinks=true,        % false: boxed links; true: colored links
  linkcolor=PineGreen,    % color of internal links
  citecolor=PineGreen,    % color of links to bibliography
  filecolor=PineGreen,    % color of file links
  urlcolor=PineGreen      % color of external links
}

\newcommand{\be}{\begin{eqnarray}}
\newcommand{\ee}{\end{eqnarray}}

\begin{document}

\title{    Multi-$\pi^+$ systems  in finite volume}

\author{Peng~Guo}
\email{pguo@csub.edu}

\affiliation{College of Physic,  Sichuan University, Chengdu, Sichuan 610065, China}
\affiliation{Department of Physics and Engineering,  California State University, Bakersfield, CA 93311, USA}
\affiliation{Kavli Institute for Theoretical Physics, University of California, Santa Barbara, CA 93106, USA}

\author{Bingwei~Long}
\email{bingwei@scu.edu.cn}
\affiliation{College of Physic,  Sichuan University, Chengdu, Sichuan 610065, China}

\date{\today}

\begin{abstract} 
 We present a formalism to describe two-$\pi^+$ and three-$\pi^+$ dynamics in finite volume,  the formalism is based on combination of a variational approach and the Faddeev method. Both pair-wise and three-body interactions are included in  the presentation.  Impacts of finite lattice spacing and the cubic lattice symmetry  are also   discussed.  To illustrate application of the formalism, the pair-wise contact interaction that  resembles the leading order  interaction terms in chiral effective theory is used  to analyze recent lattice results. 
  \end{abstract}

\maketitle

\section{Introduction}
\label{intro}

Understanding of few-hadron  interactions  is  crucial in nuclear/hadron physics.  Few-hadron dynamics provides a  
unique access to  various fundamental parameters of Quantum Chromodynamics (QCD),  for quarks and gluons only manifest themselves within hadrons due to color confinement. For instance,  the $u$- and $d$-quark mass difference  can be extracted from \mbox{$\eta \rightarrow 3 \pi$}  decay process \cite{Kambor:1995yc,Anisovich:1996tx,Schneider:2010hs,Kampf:2011wr,Guo:2015zqa,Guo:2016wsi,Colangelo:2016jmc}.
 Remarkable few-body phenomena, such as  the Efimov states  \cite{Efimov:1970zz, Braaten:2004rn} and halo nuclei  \cite{Zhukov:1993aw,Hammer:2017tjm}, have been  predicted and observed in  strong-interaction physics.   Few-body systems also offer  an outlook for many-body effects, e.g.,  three-nucleon forces \cite{Hammer:2012id,Epelbaum:2008ga,Hammer:2019poc} and fractional quantum Hall effects \cite{Fradkin:1997ge}. 

 Latest advances   have now made lattice  QCD (LQCD) a powerful quantitative tool  to study hadron physics from the first  principles.   In  recent years, realistic LQCD calculations of multi-hadron systems have been made possible \cite{Aoki:2007rd,Feng:2010es,Lang:2011mn,Aoki:2011yj,Dudek:2012gj,Dudek:2012xn,Wilson:2014cna,Wilson:2015dqa,Dudek:2016cru,Beane:2007es, Detmold:2008fn, Horz:2019rrn}. However, LQCD calculations are usually performed in a periodic box in   the Euclidean space-time, and only discrete energy  spectra are extracted from time dependent correlation functions.  That is to say, the multi-hadron dynamics is encoded in a  set of discrete energy levels in finite volume. Therefore,  mapping out  infinite-volume few-body dynamics from finite-volume energy  spectra is a key step toward understanding  multi-hadron systems from LQCD calculations.

In  the two-body sector, a  pioneering approach proposed by L\"uscher \cite{Luscher:1990ux} tends to  build connections  between infinite volume reaction amplitudes and energy levels in   a  periodic cubic box,   which was later on further extended to the cases of moving frames and coupled channels \cite{Rummukainen:1995vs,Christ:2005gi,Bernard:2008ax,He:2005ey,Lage:2009zv,Doring:2011vk,Briceno:2012yi,Hansen:2012tf,Guo:2012hv,Guo:2013vsa}.
In past few years, progresses have been made on the study of few-body systems above three-body thresholds in finite volume  \cite{Kreuzer:2008bi,Kreuzer:2009jp,Kreuzer:2012sr,Polejaeva:2012ut,Briceno:2012rv,Hansen:2014eka,Hansen:2015zga,Hansen:2016fzj,Hammer:2017uqm,Hammer:2017kms,Meissner:2014dea,Briceno:2017tce,Sharpe:2017jej,Mai:2017bge,Mai:2018djl, Doring:2018xxx, Romero-Lopez:2018rcb,Guo:2016fgl,Guo:2017ism,Guo:2017crd,Guo:2018xbv,Blanton:2019igq,Romero-Lopez:2019qrt,Blanton:2019vdk,Mai:2019fba,Guo:2018ibd,Guo:2019hih,Guo:2019ogp,Guo:2020wbl}.   To certain extent, most of these developments   may be regarded as extensions of  the  L\"uscher formula.  
L\"uscher's formula in two-body sector may   demonstrate a clear advantage: the quantization condition is given in terms of infinite volume two-body scattering amplitude or phase shift, it provides a direct  connection between infinite-volume reaction amplitudes and finite-volume  energy  spectra. Unfortunately, above three-body threshold, in addition to complication of finite volume dynamics, the infinite volume few-body reaction amplitudes usually can  not  be  easily parameterized in an analytic form, solutions of these amplitudes are given by coupled integral equations, such as Faddeev equations \cite{Faddeev:1960su,9780706505740}. Dealing  with questions regarding  infinite and finite-volume physics simultaneously  presents great challenges. 
As illustrated in Ref.~\cite{Guo:2019hih,Guo:2019ogp}, the quantization condition of few-body  systems may be presented in terms of finite-volume Green's function and  effective interaction between particles embodied by a potential.  Since infinite volume scattering amplitudes are not explicitly involved in variational approach, it may be more efficient for practical   analysis of LQCD  calculation  results. 
 
 In present work,   the variational approach  proposed in Refs. ~\cite{Guo:2018ibd,Guo:2019hih,Guo:2019ogp,Guo:2020wbl}  is applied to multi-$\pi^+$ systems. In $3\pi^+$ system, both pair-wise interactions and three-body interactions are included in our presentation.  We also take into consideration relativistic pion kinematics, the finite lattice spacing and cubic lattice symmetry effects. As our initial attempt, the  quantization conditions are  used to analyze LQCD data published in  Ref.~\cite{Horz:2019rrn} by  accounting for only pair-wise contact  interaction  in   the center-of-mass (CM) frame. The $2\pi^+$ scattering length and effective range are extracted from analysis. With  a single parameter, the coupling  constant of the pair-wise contact interaction,  we are able to make prediction on energy spectra, albeit with relatively large uncertainties.

The paper is organized as follows.  The   formalism of finite volume  multi-$\pi^+$ systems at  the continuum limit, where the lattice spacing vanishes, is   presented  in detail  in Section \ref{3bdynamics}. The  finite lattice spacing effect and cubic lattice symmetry are discussed    in Section \ref{3bdynamicsfinitea}.   Numerical results   are given  in Section  \ref{numerics}, followed by a summary in Section \ref{summary}.

\section{Relativistic  multi-$\pi^+$  systems in finite volume  at continuum limit}\label{3bdynamics}

For the future reference, we give in this section a complete presentation of our formalism that describes finite-volume dynamics of multi-$\pi^+$ systems. The formalism adopted in this work is based on the variational approach combined with the Faddeev method   which was previously discussed  in Refs.~\cite{Guo:2018ibd,Guo:2019hih,Guo:2019ogp,Guo:2020wbl}. The relativistic,  finite-volume multi-$\pi^+$ dynamics  is for the moment formulated at the continuum limit with  vanishing lattice spacing $(a=0)$, in the sense that     ultraviolet divergence is to be removed through proper regularization procedure.
The finite lattice spacing effect will be installed rather straight-forwardly later on in Section \ref{3bdynamicsfinitea}. 
In what follows, we will go through dynamical equations of the two-pion and three-pion systems, and close the section by elaborating how ultraviolet divergences are treated and how renormalization is carried out.

\subsection{Dynamical equations of $2 \pi^+$ system}\label{2pionszeroa} 
The relativistic $2 \pi^+$ system in finite volume is  governed by  the homogeneous Lippmann-Schwinger (LS) equation \cite{Guo:2019hih,Guo:2019ogp,Guo:2020wbl}:
 \begin{equation}
 \phi_{2\pi}(\mathbf{ r}) = \int_{L^3} d \mathbf{ r}' G_{2\pi}^{(\mathbf{ P})}(\mathbf{ r}-\mathbf{ r}' ; E) V (r')  \phi_{2\pi}(\mathbf{ r}'), \label{2bschrodeq}
 \end{equation}  
 where $\mathbf{ r}$ and $\mathbf{ P}$ are the relative coordinate and total momentum of the pions,  respectively. With the assumption of zero lattice spacing $(a=0)$,   $\mathbf{ r}$ is continuously distributed in finite volume, and $\int_{L^3} d \mathbf{ r}' $ stands for continuous   integration bound by  the  edges of a periodic cubic box. The spherical  two-body potential is represented by $V (r)$  where $r = |\mathbf{r}|$. The wave function for relative motion of  the two pions  satisfies periodic boundary condition  \cite{Guo:2018ibd,Guo:2019hih,Guo:2019ogp},
 \begin{equation}
 \phi_{2\pi}(\mathbf{ r} + \mathbf{ n} L)  = e^{- i \frac{\mathbf{ P}}{2} \cdot \mathbf{ n} L}  \phi_{2\pi} (\mathbf{ r}  )  , \ \  \mathbf{ n} \in \mathbb{Z}^3,
 \end{equation}
 where $L$  is the size of  the cubic lattice.  Imposing the the periodic boundary condition on the plane wave of  CM motion yields
 \begin{equation}
     \mathbf{ P} = \frac{2\pi}{L} \mathbf{d}\, , \quad \mathbf{ d} \in \mathbb{Z}^3 \, .
 \end{equation}
 We  remark that throughout  the entire paper, the explicit energy dependence  of both the wave function and the scattering amplitude  is dropped  for the convenience of presentation.   
 The two-pion Green's function is given by
 \begin{align}
 G^{(\mathbf{ P})}_{2\pi} (\mathbf{ r}  ; E)  &=    \sum_{\mathbf{ p}  } e^{i (\mathbf{ p} - \frac{\mathbf{ P}}{2}) \cdot \mathbf{ r}} \widetilde{G}^{(\mathbf{ P})}_{2\pi}  ( \mathbf{ p} ;E) , \nonumber \\
 \widetilde{G}^{(\mathbf{ P})}_{2\pi}  ( \mathbf{ p} ;E)  & =  \frac{1}{L^3} \frac{2(E_\mathbf{ p} + E_{\mathbf{ P}-\mathbf{ p}})}{2 E_\mathbf{ p} 2 E_{\mathbf{ P}-\mathbf{ p}}} \frac{1}{E^2- (E_\mathbf{ p} + E_{\mathbf{ P}-\mathbf{ p}})^2 }, \label{g2bcont}
 \end{align}
 where $E_\mathbf{ p} = \sqrt{ m_\pi^2 + \mathbf{ p}^2}$ and $\mathbf{ p} = \frac{2\pi \mathbf{ n}}{L} , \mathbf{ n} \in \mathbb{Z}^3$.

Defining finite-volume two-pion amplitude,
\begin{equation}
t^{(\mathbf{ P})}_{2\pi} (\mathbf{ k}) = - \int_{L^3} d \mathbf{ r}  e^{-i (\mathbf{ k} - \frac{\mathbf{ P}}{2}) \cdot \mathbf{ r}}  V (r )  \phi_{2\pi} (\mathbf{ r}) \, , 
\end{equation}
where  $ \mathbf{ k} = \frac{2\pi \mathbf{ n}}{L}, \mathbf{ n} \in \mathbb{Z}^3$,  one can transform Eq.~\eqref{2bschrodeq} into  the  homogeneous momentum-space LS equation:
 \begin{equation}
 t^{(\mathbf{ P})}_{2\pi}  (\mathbf{ k} )  =    \sum_{\mathbf{ p}  }    \widetilde{V}( | \mathbf{ k} - \mathbf{ p} |)   \widetilde{G}^{(\mathbf{ P})}_{2\pi}  ( \mathbf{ p} ;E) t^{(\mathbf{ P})}_{2\pi}  (\mathbf{ p}) ,  \label{2bLSeq}
 \end{equation}
 where the momentum-space potential
\begin{equation}
\widetilde{V}( k ) =  \int_{L^3} d \mathbf{ r}  e^{-i  \mathbf{ k}   \cdot \mathbf{ r}}  V (r ) \, . \label{eqn-fourierV}
\end{equation}
  According to  the variational approach \cite{Guo:2018ibd,Guo:2019hih,Guo:2019ogp,Guo:2020wbl},   the quantization condition for the two-pion system is given by
  \begin{align}
\det & \left [  \delta_{ \mathbf{ k} , \mathbf{ p}} -    \widetilde{V}(| \mathbf{ k} - \mathbf{ p}| )   \widetilde{G}^{(\mathbf{ P})}_{2\pi}  ( \mathbf{ p} ;E)   \right ] = 0, 
\nonumber \\
&\quad \quad  \quad \quad \quad \quad \quad     (\mathbf{ k} ,\mathbf{ p} ) \in \frac{2\pi \mathbf{ n}}{L} , \mathbf{ n} \in \mathbb{Z}^3 \, .
 \end{align}
 Solving for $E$ yields the discrete energy spectrum of  two interacting pions in  finite volume at the continuum limit where the lattice space approaches zero.
 
\subsection{Dynamical equations of $3 \pi^+$ system}\label{3pionszeroa} 

After  factoring out the CM motion (see Appendix \ref{nonrel3piondynamics} for the example of removing CM motion for a nonrelativistic three-particle system), the dynamics of relativistic finite-volume  three-pion system is  described by a   LS type integral equation in coordinate space,
 \begin{align}
&  \phi_{3\pi} (\mathbf{ r}_{13}, \mathbf{ r}_{23})    =\sum_{ k  =1 }^4   \int_{L^3} d \mathbf{ r}'_{13} d \mathbf{ r}'_{23}      \nonumber \\
 &\times  G_{(k)}^{(\mathbf{ P})}(\mathbf{ r}_{13}-\mathbf{ r}'_{13},\mathbf{ r}_{23}-\mathbf{ r}'_{23} ; E)    V_{(k)} (\mathbf{ r}'_{13},\mathbf{ r}'_{23} )  \phi_{3\pi} (\mathbf{ r}'_{13},\mathbf{ r}'_{23})   , \label{3bschrodeq}
 \end{align}  
 where  $\mathbf{ r}_{ij} = \mathbf{ x}_i - \mathbf{ x}_j$  is relative coordinate between the i-th and j-th pions.   $\mathbf{ r}_{13} $ and $ \mathbf{ r}_{23}$ are  chosen  to describe  the relative motion of $3\pi^+$ system.
 The pair-wise interactions between i-th and j-th identical pions are represented by
 \begin{equation}
    V_{(k)} (\mathbf{r}_{13},\mathbf{r}_{23} ) =  V (r_{ij}) \, ,
 \end{equation}
 with  $k=1,2,3$ and $k \neq i \neq j$
and $V(r)$ is the  previously defined two-body potential.  $ V_{(4)}  (\mathbf{ r}_{13},\mathbf{ r}_{23} )  $ with $k=4$   denotes the three-body force acting on all the pions.  The three-pion Green's functions are defined  as follows:
  \begin{align}
  & G^{(\mathbf{ P})}_{(k)}(\mathbf{ r}_{13}, \mathbf{ r}_{23}  ; E)  \nonumber \\
  & =      \sum_{  
              \mathbf{ p}_{1} , \mathbf{ p}_{2}    }    e^{i (\mathbf{ p}_1 - \frac{\mathbf{ P}}{2}) \cdot \mathbf{ r}_{13}}  e^{i (\mathbf{ p}_2 - \frac{\mathbf{ P}}{2}) \cdot \mathbf{ r}_{23}}  \widetilde{G}^{(\mathbf{ P})}_{(k)} ( \mathbf{ p}_1 , \mathbf{ p}_2 ;E) ,  \label{g3bcord}
  \end{align}
and 
\begin{align}
  \widetilde{G}^{(\mathbf{ P})}_{(k)} ( \mathbf{ p}_1 , \mathbf{ p}_2 ;E)  &=  2 E_{\mathbf{ p}_k}\widetilde{G}^{(\mathbf{ P})}_{3\pi}  ( \mathbf{ p}_1 , \mathbf{ p}_2 ;E),  \ \ k = 1,2,3,\nonumber \\
   \widetilde{G}^{(\mathbf{ P})}_{(4)} ( \mathbf{ p}_1 , \mathbf{ p}_2 ;E) &=  \widetilde{G}^{(\mathbf{ P})}_{3\pi}  ( \mathbf{ p}_1 , \mathbf{ p}_2 ;E), \label{g3bmom}
\end{align}           
where
   \begin{equation}
 \widetilde{G}^{(\mathbf{ P})}_{3\pi}  ( \mathbf{ p}_1 ,  \mathbf{ p}_2 ;E)   =\frac{1}{L^6}    \frac{2 \sum_{i=1}^3 E_{\mathbf{ p}_i }  }{2 E_{\mathbf{ p}_1  } 2 E_{\mathbf{ p}_2  } 2 E_{\mathbf{ p}_3  }}   \frac{1}{E^2-  ( \sum_{i=1}^3 E_{\mathbf{ p}_i }  )^2 }\, .\label{g3bcont}
 \end{equation}
 The total momentum
\begin{equation}
\mathbf{ P} =  \mathbf{ p}_1 + \mathbf{ p}_2 + \mathbf{ p}_3 = \frac{2\pi}{L} \mathbf{ d}  \, , \ \  \mathbf{ d}  \in \mathbb{Z}^3 \, , 
\end{equation}
  where
\begin{equation}
\mathbf{ p}_{i} = \frac{2\pi \mathbf{ n}_{i}}{L}, \mathbf{ n}_{i} \in \mathbb{Z}^3 \, ,
\end{equation}
is the  momentum of the $i$-th particle.
  The relativistic kinematic  factors, $2 E_{\mathbf{ p}_k}$ for $k=1,2,3$, in Eq.(\ref{g3bmom}) are  associated  with the relativistic normalization of  the free propagating spectator particle in presence of the pair-wise interaction between  the i-th and j-th particles. The  relativistic LS equation  may be  derived from  the Bethe-Salpeter equation,  see  Appendix \ref{reductionBS}. The   three-body wave function must satisfy the periodic boundary condition \cite{Guo:2019ogp}:
 \begin{equation}
 \phi_{3\pi}  (\mathbf{ r}_{13} + \mathbf{ n}_{1} L, \mathbf{ r}_{23} + \mathbf{ n}_2 L)  = e^{- i \frac{\mathbf{ P}}{3} \cdot  ( \mathbf{ n}_1 L +  \mathbf{ n}_2 L )}  \phi_{3\pi} (\mathbf{ r}_{13}, \mathbf{ r}_{23})  ,
 \end{equation}
where $\mathbf{ n}_{1,2} \in \mathbb{Z}^3$.

 As suggested in  Refs.~\cite{Guo:2018ibd,Guo:2019hih,Guo:2019ogp,Guo:2020wbl},  the Faddeev amplitudes may be introduced by 
 \begin{align}
 t_{(k)}^{(\mathbf{ P})}(\mathbf{ k}_1, \mathbf{ k}_2) = - & \int_{L^3} d \mathbf{ r}_{13}  d \mathbf{ r}_{23}  e^{-i (\mathbf{ k}_1 - \frac{\mathbf{ P}}{3}) \cdot \mathbf{ r}_{13} } e^{-i (\mathbf{ k}_2 - \frac{\mathbf{ P}}{3}) \cdot \mathbf{ r}_{23} }  \nonumber \\
 & \times  V (r_{ij} )  \phi_{3\pi} (\mathbf{ r}_{13}, \mathbf{ r}_{23}) ,  \ \ \ k \neq i \neq j , \nonumber \\
  t_{(4)}^{(\mathbf{ P})}(\mathbf{ k}_1, \mathbf{ k}_2) = - & \int_{L^3} d \mathbf{ r}_{13}  d \mathbf{ r}_{23}  e^{-i (\mathbf{ k}_1 - \frac{\mathbf{ P}}{3}) \cdot \mathbf{ r}_{13} } e^{-i (\mathbf{ k}_2 - \frac{\mathbf{ P}}{3}) \cdot \mathbf{ r}_{23} }  \nonumber \\
 & \times  V_{(4)} (\mathbf{ r}_{13} ,\mathbf{ r}_{23} )  \phi_{3\pi} (\mathbf{ r}_{13}, \mathbf{ r}_{23}) . \label{tkdef}
 \end{align}
   Equation~\eqref{3bschrodeq} is thus turned into coupled equations:
    \begin{align}
& t_{(k)}^{(\mathbf{ P})}(\mathbf{ k}_1, \mathbf{ k}_2)    =        \sum_{ 
              \mathbf{ p}_{1}   , \mathbf{ p}_{2}    }     \widetilde{V}_{(k)} ( \mathbf{ k}_1 - \mathbf{ p}_1,\mathbf{ k}_2 -  \mathbf{ p}_2 )     \nonumber \\
 &\times  \widetilde{G}^{(\mathbf{ P})}_{3\pi}  ( \mathbf{ p}_1 ,  \mathbf{ p}_2 ;E)    \left [ \sum_{k'  =1 }^3  2 E_{\mathbf{ p}_{k'}}   t_{(k')}^{(\mathbf{ P})}(\mathbf{ p}_1, \mathbf{ p}_2)  +   t_{(4)}^{(\mathbf{ P})}(\mathbf{ p}_1, \mathbf{ p}_2) \right ]  , \label{3bLSoriginal}
 \end{align}  
 where $\widetilde{V}_{(k )} $'s are the Fourier transform of pair-wise and three-body interaction potentials,
 \begin{align}
 & \widetilde{V}_{(k )}  ( \mathbf{ k}_1  ,\mathbf{ k}_2   ) =     \int_{L^3} d \mathbf{ r}_{13}  d \mathbf{ r}_{23}   e^{ - i  \mathbf{ k}_1   \cdot  \mathbf{ r}_{13}   }  e^{ - i   \mathbf{ k}_2  \cdot   \mathbf{ r}_{23} } V (r_{ij} )|_{k =1,2,3}^{k\neq i \neq j},   \nonumber \\
 & \widetilde{V}_{(4)}  ( \mathbf{ k}_1  ,\mathbf{ k}_2   )  =     \int_{L^3} d \mathbf{ r}_{13}  d \mathbf{ r}_{23}   e^{ - i  \mathbf{ k}_1   \cdot  \mathbf{ r}_{13}   }  e^{ - i   \mathbf{ k}_2  \cdot   \mathbf{ r}_{23} } V_{(4)} (\mathbf{ r}_{13} ,\mathbf{ r}_{23} )  .
 \end{align}
Two-body interactions $\widetilde{V}_{(1,2,3) }$ are related to $\widetilde{V}(k)$, as defined in Eq.~\eqref{eqn-fourierV}, through      
 \begin{align}
 \widetilde{V}_{(2)}  ( \mathbf{ k}_1  ,\mathbf{ k}_2   ) &=    \widetilde{V}   ( k_1   )  L^3 \delta_{\mathbf{ k}_2, \mathbf{ 0}} ,     \nonumber \\
  \widetilde{V}_{(1)}  ( \mathbf{ k}_1  ,\mathbf{ k}_2   ) &=      \widetilde{V}   ( k_2   )  L^3 \delta_{\mathbf{ k}_1, \mathbf{ 0}}  ,    \nonumber \\ 
   \widetilde{V}_{(3)}  ( \mathbf{ k}_1  ,\mathbf{ k}_2   ) &=    \widetilde{V}   (    \frac{|  \mathbf{ k}_1 -\mathbf{ k}_2 |  }{2}    )  L^3 \delta_{ \mathbf{ k}_1 , - \mathbf{ k}_2   }  ,
 \end{align}
 where $\delta_{\mathbf{ k}, \mathbf{ p}} $  denotes the 3D Kronecker delta function.

\subsubsection{Exchange symmetry of $3\pi^+$ system}\label{exchangesym} 
 As is the case for any identical bosons, the wave function of $3\pi^+$ system must be invariant under   exchange of  any pair of pions, for example,
 \begin{equation}
  \phi_{3\pi}  (\mathbf{ r}_{13}, \mathbf{ r}_{23})  \stackrel{   1 \leftrightarrow 3 }{ = } \phi_{3\pi}  (\mathbf{ r}_{31}, \mathbf{ r}_{21})  \stackrel{   2 \leftrightarrow 1 }{ = }   \phi_{3\pi}  (\mathbf{ r}_{12}, \mathbf{ r}_{32}), \label{symwav}
 \end{equation}
 and $\phi_{3\pi}  (\mathbf{ r}_{ik}, \mathbf{ r}_{jk})  \stackrel{   i \leftrightarrow j }{ = }  \phi_{3\pi}  ( \mathbf{ r}_{jk},\mathbf{ r}_{ik})  $. The exchange symmetry of  three-body wave function suggests that   $t_{(1,2,3)}^{(\mathbf{ P})}$ are related:  
  \begin{align}
   t_{(1)}^{(\mathbf{ P})}(\mathbf{ k}_1, \mathbf{ k}_2)    & =  t^{(\mathbf{ P})}_{(2)} (\mathbf{ k}_2, \mathbf{ k}_1), \nonumber \\
   t_{(3)}^{(\mathbf{ P})}(\mathbf{ k}_1, \mathbf{ k}_2) &=  t^{(\mathbf{ P})}_{(2)} (\mathbf{ k}_1, \mathbf{ k}_3  )        =  t^{(\mathbf{ P})}_{(2)} (\mathbf{ k}_2, \mathbf{ k}_3 ) \, ,
 \end{align}   
  where 
  $\mathbf{ P} = \mathbf{ k}_1  + \mathbf{ k}_2 +  \mathbf{ k}_3$.

Using the definition of  $t_{(2)}^{(\mathbf{ P})}$ in Eq.(\ref{tkdef}) and symmetry relations of wave function in Eq.(\ref{symwav}), we  find useful symmetry  properties of  $t^{(\mathbf{ P})}_{(2)}  $:
  \begin{equation}
    t^{(\mathbf{ P})}_{(2)} (  \mathbf{ k}_3, \mathbf{ k}_2)   =  t^{(\mathbf{ P})}_{(2)} (  \mathbf{ k}_1 ,   \mathbf{ k}_2) ,   \ \   t^{(\mathbf{ P})}_{(2)} (  \mathbf{ k}_3, \mathbf{ k}_1)   =  t^{(\mathbf{ P})}_{(2)} (  \mathbf{ k}_2 ,   \mathbf{ k}_1) .
 \end{equation}
 In addition to that of the wave function, exchange symmetry of three-body potential $V_{(4)}$ also constrains the amplitude  $t_{(4)}$:
   \begin{equation}
    t_{(4)}^{(\mathbf{ P})}(  \mathbf{ k}_1 , \mathbf{ k}_2)   =  t_{(4)}^{(\mathbf{ P})}(  \mathbf{ k}_2 , \mathbf{ k}_1) =  t_{(4)}^{(\mathbf{ P})}( \mathbf{ k}_1, \mathbf{ k}_3  )    = \cdots  .
 \end{equation}

  \begin{figure}
\begin{center}
\includegraphics[width=0.48\textwidth]{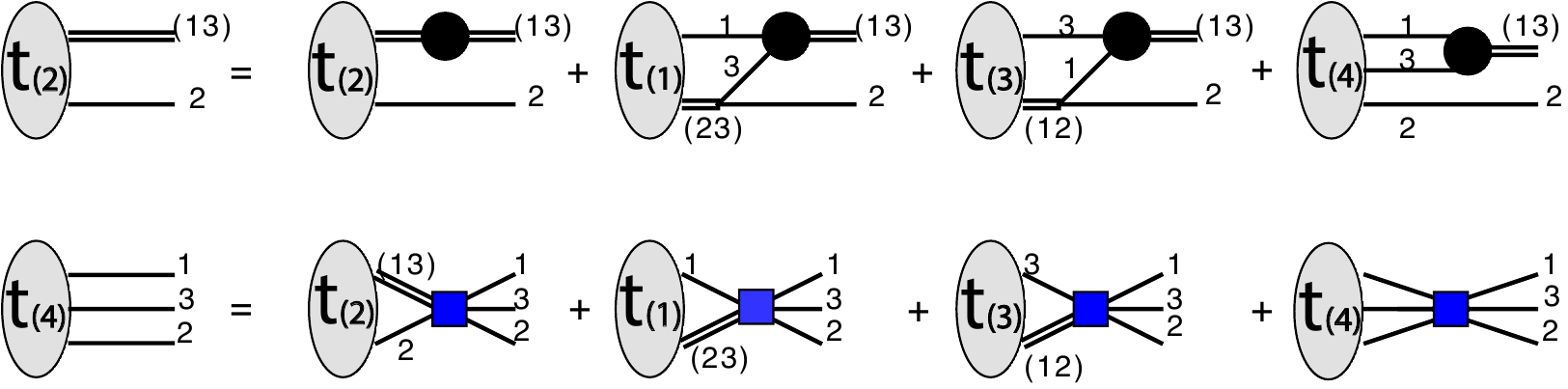}
\caption{  Diagrammatic representation of Eq.(\ref{3bLSred1}) and Eq.(\ref{3bLSred2}), pair-wise and three-body interactions are represented by black solid circle and blue solid square.   }\label{feynmanplot}
\end{center}
\end{figure}

\subsubsection{Further reduction of $3\pi^+$ LS equations} 

Using  the aforementioned symmetry relations of Faddeev amplitudes,   Eq.(\ref{3bLSoriginal})   are further reduced into  the following equations:
  \begin{align}
& t_{(2)}^{(\mathbf{ P})}(\mathbf{ k}_1, \mathbf{ k}_2)    =        \sum_{ 
              \mathbf{ p}_{1}    }      \widetilde{V}   ( | \mathbf{ k}_1- \mathbf{ p}_1   |) L^3  \widetilde{G}^{(\mathbf{ P})}_{3\pi}  ( \mathbf{ p}_1 ,  \mathbf{ k}_2 ;E)   \nonumber \\
 &\times     \bigg [ 2 E_{\mathbf{ k}_{2}}   t_{(2)}^{(\mathbf{ P})}(\mathbf{ p}_1, \mathbf{ k}_2) + 2 E_{\mathbf{ p}_{1}}   t_{(2)}^{(\mathbf{ P})}( \mathbf{ k}_2,\mathbf{ p}_1)  \nonumber \\
 & \quad    +2 E_{\mathbf{ P} - \mathbf{ p}_1- \mathbf{ k}_{2}}   t_{(2)}^{(\mathbf{ P})}(\mathbf{ p}_1, \mathbf{ P}-\mathbf{ p}_1- \mathbf{ k}_2) +   t_{(4)}^{(\mathbf{ P})}(\mathbf{ p}_1, \mathbf{ k}_2) \bigg ]  , \label{3bLSred1}
 \end{align}  
 and
   \begin{align}
 &  t_{(4)}^{(\mathbf{ P})}  (\mathbf{ k}_1, \mathbf{ k}_2)  =        \sum_{ 
              \mathbf{ p}_{1}   , \mathbf{ p}_{2}    }     \widetilde{V}_{(4)} ( \mathbf{ k}_1 - \mathbf{ p}_1,\mathbf{ k}_2 -  \mathbf{ p}_2 )    \widetilde{G}^{(\mathbf{ P})}_{3\pi}  ( \mathbf{ p}_1 ,  \mathbf{ p}_2 ;E)    \nonumber \\
 &\times     \bigg [   2 E_{\mathbf{ p}_{2}}   t_{(2)}^{(\mathbf{ P})}(\mathbf{ p}_1, \mathbf{ p}_2) + 2 E_{\mathbf{ p}_{1}}   t_{(2)}^{(\mathbf{ P})}(\mathbf{ p}_2, \mathbf{ p}_1)   \nonumber \\
 & \quad + 2 E_{\mathbf{ P} - \mathbf{ p}_1- \mathbf{ p}_2 }   t_{(2)}^{(\mathbf{ P})}(\mathbf{ p}_1, \mathbf{ P} - \mathbf{ p}_1- \mathbf{ p}_2)    +   t_{(4)}^{(\mathbf{ P})}(\mathbf{ p}_1, \mathbf{ p}_2) \bigg ]  .\label{3bLSred2}
 \end{align}
 The diagrammatic representations of  Eq.(\ref{3bLSred1}) and Eq.(\ref{3bLSred2})  are  shown in Fig.~\ref{feynmanplot}. 
It may be convenient to consolidate Eqs.(\ref{3bLSred1}) and (\ref{3bLSred2})  in  the matrix form:
  \begin{align}
&\begin{bmatrix} 
t^{(\mathbf{ P})}_{(2)} (\mathbf{ k}_1, \mathbf{ k}_2)   \\ 
t_{(4)}^{(\mathbf{ P})}(\mathbf{ k}_1, \mathbf{ k}_2) 
\end{bmatrix} =       \sum_{ 
              \mathbf{ p}_{1}   , \mathbf{ p}_{2}    }   \widetilde{G}^{(\mathbf{ P})}_{3\pi}  ( \mathbf{ p}_1 ,  \mathbf{ p}_2 ;E)   \nonumber \\
 &  \quad    \times \mathcal{K}  ( \mathbf{ k}_1, \mathbf{ k}_2; \mathbf{ p}_1, \mathbf{ p}_2 )  \begin{bmatrix} 
t^{(\mathbf{ P})}_{(2)} (\mathbf{ p}_1, \mathbf{ p}_2)   \\ 
t_{(4)}^{(\mathbf{ P})}(\mathbf{ p}_1, \mathbf{ p}_2) 
\end{bmatrix}     , \label{3bLSeq}
 \end{align}  
 where  $( \mathbf{ k}_i, \mathbf{ p}_{i} ) \in \frac{2\pi \mathbf{ n}}{L}  ,
             \mathbf{ n}  \in \mathbb{Z}^3$.   Elements of the matrix function $ \mathcal{K}  ( \mathbf{ k}_1, \mathbf{ k}_2; \mathbf{ p}_1, \mathbf{ p}_2 ) $   are given by
 \begin{align}
 & \mathcal{K}_{11}  ( \mathbf{ k}_1, \mathbf{ k}_2; \mathbf{ p}_1, \mathbf{ p}_2 )   =  2 E_{\mathbf{ p}_{2} }  L^3\bigg [    \delta_{\mathbf{ p}_2, \mathbf{ k}_2 }    \widetilde{V}   ( | \mathbf{ k}_1- \mathbf{ p}_1 |  )   \nonumber \\
 & \quad \quad  +  \delta_{\mathbf{ p}_1, \mathbf{ k}_2 }  \left (     \widetilde{V}   ( | \mathbf{ k}_1- \mathbf{ p}_2 |  )   +    \widetilde{V}   (| \mathbf{ k}_1  -  \mathbf{ p}_3 | )      \right )    \bigg ]  ,  \nonumber \\
& \mathcal{K}_{12}  ( \mathbf{ k}_1, \mathbf{ k}_2; \mathbf{ p}_1, \mathbf{ p}_2 )   =   L^3  \delta_{\mathbf{ p}_2, \mathbf{ k}_2 }  \widetilde{V}   ( | \mathbf{ k}_1- \mathbf{ p}_1 |  )      ,  \nonumber \\
&  \mathcal{K}_{21}  ( \mathbf{ k}_1, \mathbf{ k}_2; \mathbf{ p}_1, \mathbf{ p}_2 )    = 2 E_{\mathbf{ p}_{2}}       \bigg [    \widetilde{V}_{(4)} ( \mathbf{ k}_1 - \mathbf{ p}_1,\mathbf{ k}_2 -  \mathbf{ p}_2 )     \nonumber \\
&\quad \quad +    \widetilde{V}_{(4)} ( \mathbf{ k}_1 - \mathbf{ p}_2,\mathbf{ k}_2 -  \mathbf{ p}_1 )  +  \widetilde{V}_{(4)} ( \mathbf{ k}_1 - \mathbf{ p}_3 ,\mathbf{ k}_2 -  \mathbf{ p}_1 )    \bigg ]     , \nonumber \\
&  \mathcal{K}_{22}  ( \mathbf{ k}_1, \mathbf{ k}_2; \mathbf{ p}_1, \mathbf{ p}_2 ) =     \widetilde{V}_{(4)} ( \mathbf{ k}_1 - \mathbf{ p}_1,\mathbf{ k}_2 -  \mathbf{ p}_2 )      \, .
 \end{align}
    The coupled equations,  Eqs.(\ref{3bLSred1}) and (\ref{3bLSred2}),   or  the matrix form of them   Eq.(\ref{3bLSeq}), will serve as basic dynamical equations for $3\pi^+$ system.   The entire   $3\pi^+$ energy spectrum in a cubic box can   be produced by the quantization condition  from Eq.(\ref{3bLSeq}):  
 \begin{equation}
 \det \left [ \mathbb{I} -      \widetilde{G}^{(\mathbf{ P})}  \mathcal{K}   \right ] =0   .
 \end{equation}

More technical aspects need to be spelled out  in order to apply the above quantization condition: renormalization procedure, projection of the spectrum according to irreducible representations of cubic symmetry group, and finite lattice spacing effect. All these mentioned factors hence ultimately will add extra layer of technical complication on top of  Eq.(\ref{3bLSeq}).  As a simple illustration of our formalism, a specific choice about interactions is made in this work: the two-body potential is the zero-range potential and the three-body interaction is turned off. The contact interaction resembles  the leading order  terms of the chiral Lagrangian \cite{Hammer:2017uqm,Hammer:2017kms,Rusetsky:2019gyk}.  With contact interaction, the $3\pi^+$ LS equation is simplified  considerably.  We now turn to renormalization of the multi-$\pi^+$ LS equations.

\subsection{Pair-wise contact interaction and renormalization}\label{contactpot}

Since contact interactions allow particles to get arbitrarily close to each other, it is often necessary to regularize the ultraviolet part of the dynamics. This is indeed the case with our illustrative choice of multiple-pion interactions. We will discuss renormalization of both $2\pi^+$ and $3\pi^+$ LS equations in what follows.

\subsubsection{Renormalization of $2\pi^+$ LS equation}

\begin{figure}
\begin{center}
\includegraphics[width=0.45\textwidth]{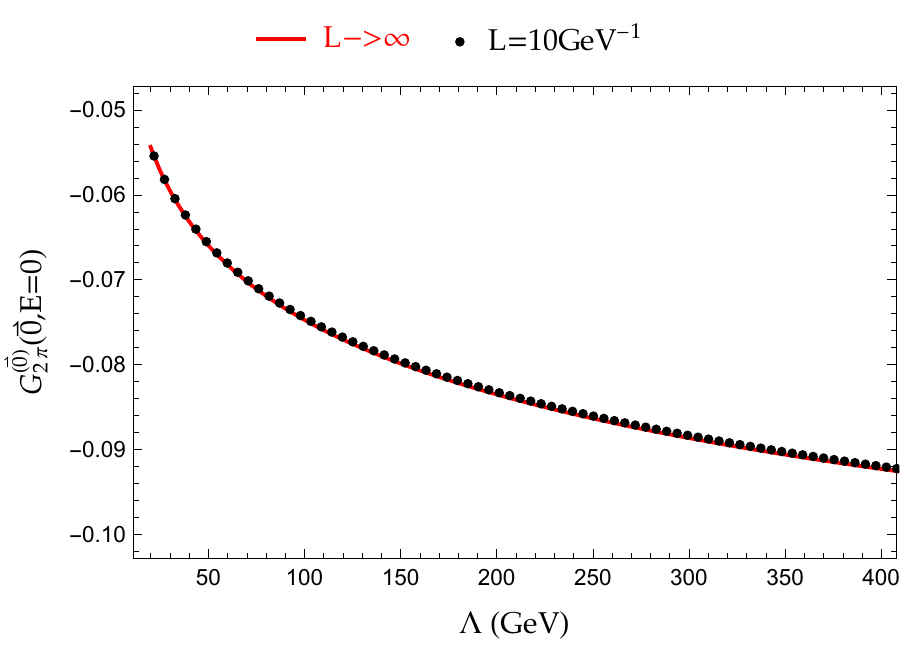}
\caption{ Plot of  cufoff dependence of   $\sum_{\mathbf{ p}  }^{    | \mathbf{ p}| < \Lambda  }       \widetilde{G}^{(\mathbf{ 0})}_{2\pi} ( \mathbf{ p} ;0)$ (black dots) vs.  $\frac{1}{8\pi^2} (1- \ln \frac{2\Lambda}{m_\pi})  $ (red curve), where   $L=10 \mbox{GeV}^{-1}$ and $m_\pi=0.200 \mbox{GeV}$.  The black dots are shifted by a constant value:  $ \sum_{\mathbf{ p}  }^{    | \mathbf{ p}| < \Lambda_0  }       \widetilde{G}^{(\mathbf{ 0})}_{2\pi} ( \mathbf{ p} ;0) - \frac{1 }{8\pi^2} (1- \ln \frac{2\Lambda_0}{m_\pi}) $, where $\Lambda_0 = \frac{40 \pi \sqrt{3} }{L}   $,   so that black dots and red curve overlap when  $\Lambda = \Lambda_0$. }\label{G2bcutoffplot}
\end{center}
\end{figure}

The zero-range potential, $V(r)= V_0 \delta (\mathbf{ r})$   and $\widetilde{V}( k  )  = V_0$, is the simplest case of separable potentials. More specifically in this case, 
\begin{equation}
t^{(\mathbf{ P})}_{2\pi} (\mathbf{ k}) \equiv t^{(\mathbf{ P})}_{2\pi}   
\end{equation}
 is independent of $\mathbf{k}$  for fixed $E$, and  is factored out of the two-body homogeneous LS equation \eqref{2bLSeq}:
 \begin{equation}
 1=  V_0  \sum_{\mathbf{ p}  }     \widetilde{G}^{(\mathbf{ P})}_{2\pi} ( \mathbf{ p} ;E)   .  \label{2bqcunrenorm}
 \end{equation}
The infinite momentum sum of  two-pion Green's function \eqref{g2bcont} is  divergent, and the divergence  can be regularized by  imposing a sharp ultraviolet momentum cutoff $\Lambda$. 
It is instructive to display the divergence  in infinite volume and at the continuum limit: 
  \begin{align}
  \sum_{\mathbf{ p}}^{    | \mathbf{ p}| < \Lambda  }     \widetilde{G}^{(\mathbf{ 0})}_{2\pi} ( \mathbf{ p} ;0) &  \stackrel{L \rightarrow \infty}{  \rightarrow}  - \frac{1}{8 \pi^2} \int^{\Lambda} d p \frac{p^2}{E_\mathbf{ p}^3}  \nonumber \\
  &  \stackrel{\Lambda \rightarrow \infty}{\rightarrow}   - \frac{1}{8\pi^2} \ln \frac{\Lambda}{m_\pi} + \frac{1-\ln 2}{8\pi^2}\, .
 \end{align}
 For finite values of $\mathbf{ P}$ and $E$, the divergence remains same, or equivalently, adds corrections in powers of $P/\Lambda$ and $\sqrt{m_\pi E}/\Lambda$.
 The cutoff dependence of 
 $$
 \sum_{\mathbf{ p}  }^{    | \mathbf{ p}| < \Lambda  }       \widetilde{G}^{(\mathbf{ P})}_{2\pi} ( \mathbf{ p} ;E = 0)
 $$
  compared with its infinite-volume counterpart, 
$$
- \frac{1}{8 \pi^2} \int^{\Lambda} d p \frac{p^2}{E_\mathbf{ p}^3}    \stackrel{\Lambda \rightarrow \infty}{\rightarrow}     \frac{1}{8\pi^2}(1- \ln \frac{2 \Lambda}{m_\pi} )   \,  ,
$$
as $\Lambda \rightarrow \infty$, is  shown in Fig.~\ref{G2bcutoffplot}.

 Finite lattice spacing  provides a natural  regularization on ultraviolet divergence.  The  physical observables   must not depend on the cutoff or choice of lattice spacing,  and this is to be assured by renormalization procedure. The bare interaction strength, $V_0$, must be redefined to absorb ultraviolet divergence of the momentum sum of Green's function by
 \begin{equation}
 \frac{1}{V_0} = \frac{1}{V_R (\mu) } + \sum_{\mathbf{ p}  }^{    | \mathbf{ p}| < \Lambda  }       \widetilde{G}^{(\mathbf{ 0})}_{2\pi} ( \mathbf{ p} ;\mu) \, , \label{V0renorm}
 \end{equation}
where $V_R (\mu)$ stands for the renormalized physical coupling strength  at the renormalization scale, $\mu$. 
The two-body quantization condition (\ref{2bqcunrenorm}) can be rewritten with the renormalized coupling as
 \begin{equation}
 \frac{1}{V_R (\mu)}  =    \sum_{\mathbf{ p}  }^{    | \mathbf{ p}| < \Lambda  }       \widetilde{G}^{(\mathbf{ P})}_{2\pi} ( \mathbf{ p} ;E)  - \sum_{\mathbf{ p}  }^{    | \mathbf{ p}| < \Lambda  }       \widetilde{G}^{(\mathbf{ 0})}_{2\pi} ( \mathbf{ p} ;\mu) \, ,  \label{2bqcrenorm}
 \end{equation}
where the cutoff  dependencies from the summations on the right-hand side cancel out at the limit $\Lambda \to \infty$, so 
ultraviolet divergence is now removed from the quantization condition.

  \begin{figure}
\begin{center}
\includegraphics[width=0.45\textwidth]{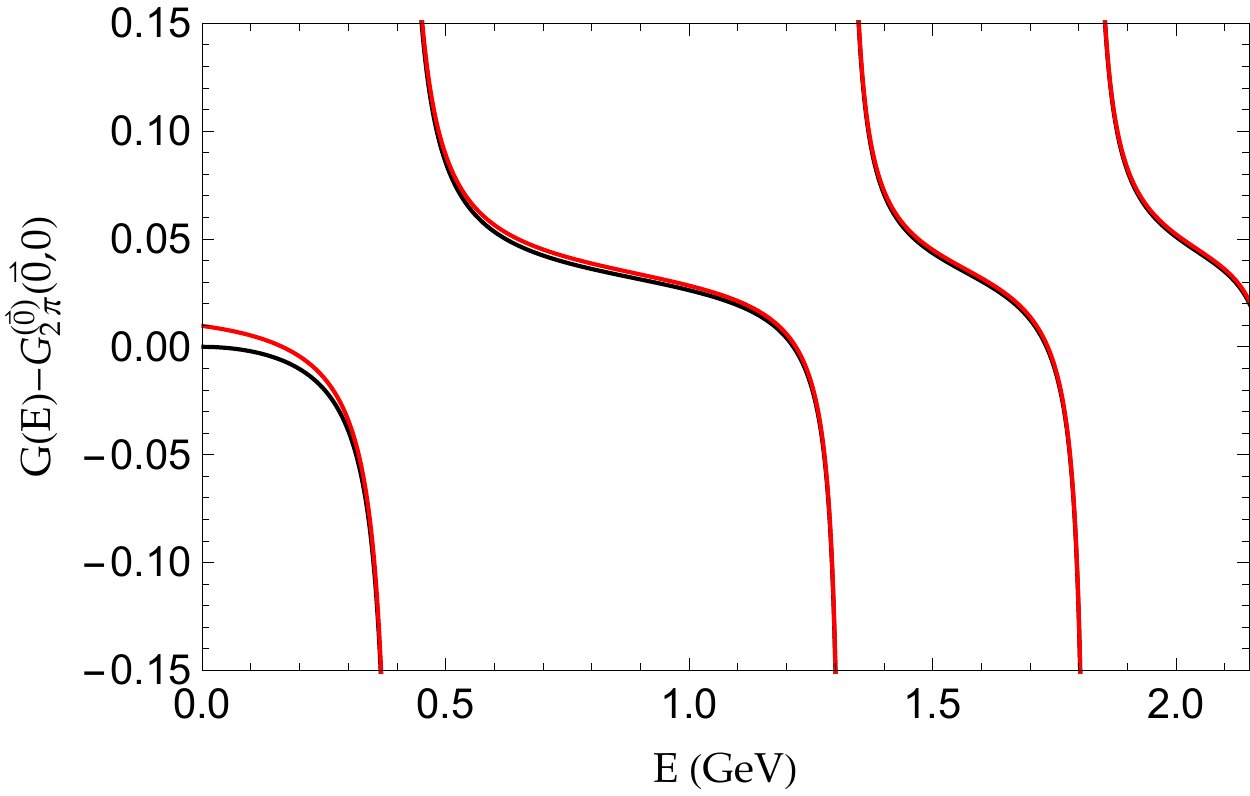}
\caption{ Plot of     $\sum_{\mathbf{ p}  }^{    | \mathbf{ p}| < \Lambda  }    \left [    \widetilde{G}^{(\mathbf{ 0})}_{2\pi} ( \mathbf{ p} ;E)  -    \widetilde{G}^{(\mathbf{ 0})}_{2\pi} ( \mathbf{ p} ;\mu) \right ]$ (black curve) vs.    $\sum_{\mathbf{ p}  }^{    | \mathbf{ p}| < \Lambda  }    \left [  2 E_\mathbf{ 0} L^3  \widetilde{G}^{(\mathbf{ 0})}_{3\pi} ( \mathbf{ p}, \mathbf{ 0} ;E+2 m_\pi)  -    \widetilde{G}^{(\mathbf{ 0})}_{2\pi} ( \mathbf{ p} ;\mu) \right ]$ (red curve)  with $\mu =0 \mbox{GeV}$, $L=10 \mbox{GeV}^{-1}$ and $m_\pi=0.200 \mbox{GeV}$. }\label{G2bvsG3bplot}
\end{center}
\end{figure}

\subsubsection{Renormalization of $3\pi^+$ LS equation}

With three-body interaction $V_{(4)}$ turned off, $t^{(\mathbf{ P})}_{(4)}$ vanishes in Eqs.~\eqref{3bLSred1} and \eqref{3bLSred2}.  Due to  the zero-range nature of the interaction, the amplitude $t^{(\mathbf{ P})}_{(2)}(\mathbf{ k}_1, \mathbf{ k}_2)  $  associated with   two-pion interaction between  the pair $(13)$ depends only on momentum $\mathbf{ k}_2$ . So it is  appropriate to rename  $t^{(\mathbf{ P})}_{(2)}(\mathbf{ k}_1, \mathbf{ k}_2)  $  as
\begin{equation}
  t^{(\mathbf{ P})}_{3\pi}(  \mathbf{ k}_2) =t^{(\mathbf{ P})}_{(2)}(\mathbf{ k}_1, \mathbf{ k}_2).
\end{equation}
The $3 \pi^+$ LS equation \eqref{3bLSeq} is therefore reduced to  
  \begin{align}
& \left [ 1 -  V_0   \sum_{ 
              \mathbf{ p}    }   2 E_{\mathbf{ k}_2 }    L^3  \widetilde{G}^{(\mathbf{ P})}_{3\pi} ( \mathbf{ p} ,  \mathbf{ k}_2 ;E)   \right ]   t^{(\mathbf{ P})}_{3\pi}( \mathbf{ k}_2)   \nonumber \\
              &=          2 V_0 \sum_{ 
              \mathbf{ p}    }     2 E_{\mathbf{ p}}    L^3  \widetilde{G}^{(\mathbf{ P})}_{3\pi} ( \mathbf{ p} ,  \mathbf{ k}_2 ;E)   t^{(\mathbf{ P})}_{3\pi}(  \mathbf{ p})    . \label{3bLScontact}
 \end{align}  
The first line of Eq.(\ref{3bLScontact})  describes interaction  within the pair $(13)$,  with the second pion acting as the spectator. The second line of Eq.(\ref{3bLScontact})  represents crossed-channel interactions through exchanging the second particle between the pairs. The direct-channel terms  in Eq.(\ref{3bLScontact}) resemble the leading order isobar contribution in  Khuri-Treiman approach \cite{Khuri:1960zz,Bronzan:1963mby,Aitchison:1965zz,Guo:2014vya,Guo:2014mpp,Danilkin:2014cra,Guo:2015kla}.
The crossed-channel terms are  associated with rescattering   corrections from other pairs  into isobar pair $(13)$.    
It can be illustrated  quite straightforwardly by iterations of Eq.(\ref{3bLScontact}) that crossed-channel contributions are  not UV divergent, and that the UV divergence emerge only in direct-channel term.  
So, as far as renormalization is concerned, crossed-channel interactions can be ``turned off''. We arrive at an equation similar to Eq.~\eqref{2bqcunrenorm}:
\begin{equation}
1 =  V_0   \sum_{ 
              \mathbf{ p}    }   2 E_{\mathbf{ k}_2 }    L^3  \widetilde{G}^{(\mathbf{ P})}_{3\pi} ( \mathbf{ p} ,  \mathbf{ k}_2 ;E) \, , \label{3bLSV13}
\end{equation}
where $\mathbf{ k}_2$ is momentum of the spectator pion.
 Equations ~\eqref{3bLSV13} and \eqref{2bqcunrenorm} are not identical due to the relativistic kinematics  brought by the spectator particle on top of the pair $(13)$:
 \begin{align}
&    2 E_{\mathbf{ k}_2 }    L^3  \widetilde{G}^{(\mathbf{ P})}_{3\pi} ( \mathbf{ k}_1 ,  \mathbf{ k}_2 ;E)  \nonumber \\
& =   \frac{1}{L^3}    \frac{2  (E_{\mathbf{ k}_1 }  +  E_{\mathbf{ k}_3 } + E_{\mathbf{ k}_2} )  }{2 E_{\mathbf{ k}_1  }  2 E_{\mathbf{ k}_3  }}   \frac{1}{E^2-  (   E_{\mathbf{ k}_1 } +  E_{\mathbf{ k}_3 }   + E_{\mathbf{ k}_2} )^2 }  ,
\end{align}
 compared to 
  \begin{align}
    \widetilde{G}^{(\mathbf{ P})}_{2\pi} ( \mathbf{ k}_1   ;E) =   \frac{1}{L^3}     \frac{2  (E_{\mathbf{ k}_1 } + E_{\mathbf{ k}_3}   )  }{2 E_{\mathbf{ k}_1  }  2 E_{\mathbf{ k}_3  }}   \frac{1}{E^2-  (   E_{\mathbf{ k}_1 } + E_{\mathbf{ k}_3}   )^2 }  ,
\end{align}
 where $\mathbf{ P} = \mathbf{ k}_1 + \mathbf{ k}_3  $  is the total momentum of the pair $(13)$.  Nonetheless, the multi-particle Green's function is dominated by the   location of   poles of particles propagator,  so the dominant contributions of the $2\pi^+$ and $3\pi^+$ Green's functions behave in a similar way,
 \begin{align}
&    2 E_{\mathbf{ k}_2 }    L^3  \widetilde{G}^{(\mathbf{ P})}_{3\pi} ( \mathbf{ k}_1 ,  \mathbf{ k}_2 ;E)  \nonumber \\
& \sim  \frac{1}{L^3}    \frac{1 }{2 E_{\mathbf{ k}_1  }  2 E_{\mathbf{ k}_3  }}   \frac{1}{E-  (   E_{\mathbf{ k}_1 } +  E_{\mathbf{ k}_3 }   + E_{\mathbf{ k}_2} ) }  ,
\end{align}
and 
  \begin{align}
    \widetilde{G}^{(\mathbf{ P})}_{2\pi} ( \mathbf{ k}_1   ;E) \sim  \frac{1}{L^3}     \frac{1 }{2 E_{\mathbf{ k}_1  }  2 E_{\mathbf{ k}_3  }}   \frac{1}{E-  (   E_{\mathbf{ k}_1 } + E_{\mathbf{ k}_3}   ) }  .
\end{align}
Therefore, the asymptotic high energy behavior of infinite momentum sum in   Eq.(\ref{3bLSV13}) and Eq.(\ref{2bqcunrenorm}) should be exactly same, with a momentum cutoff,  
 \begin{equation}
    \sum_{ 
              \mathbf{ p}     }^{    | \mathbf{ p}| < \Lambda  }     2 E_{\mathbf{ k}_2 }    L^3  \widetilde{G}^{(\mathbf{ P})}_{3\pi} ( \mathbf{ p} ,  \mathbf{ k}_2 ;0)   
               \sim   -     \frac{1}{8\pi^2} \ln \frac{\Lambda}{m_\pi}  +   \mbox{finite part} .
 \end{equation}  
The renormalization procedure in $3\pi^+$ LS equation  thus can be carried out in the same way as in the $2\pi^+$ sector, see Fig.~\ref{G2bvsG3bplot} for the comparison of
 $$
 \sum_{\mathbf{ p}  }^{    | \mathbf{ p}| < \Lambda  }    \left [    \widetilde{G}^{(\mathbf{ 0})}_{2\pi} ( \mathbf{ p} ;E)  -    \widetilde{G}^{(\mathbf{ 0})}_{2\pi} ( \mathbf{ p} ;\mu) \right ]
 $$
 and
 $$\sum_{\mathbf{ p}  }^{    | \mathbf{ p}| < \Lambda  }    \left [  2 E_\mathbf{ 0} L^3  \widetilde{G}^{(\mathbf{ 0})}_{3\pi} ( \mathbf{ p}, \mathbf{ 0} ;E+2 m_\pi)  -    \widetilde{G}^{(\mathbf{ 0})}_{2\pi} ( \mathbf{ p} ;\mu) \right ] \, .
 $$

Using Eq.(\ref{V0renorm}) and redefining the bare coupling $V_0$, 
the renormalized $3\pi^+$ LS equation can be given in a compact form:
 \begin{align}
   t^{(\mathbf{ P})}_{3\pi}( \mathbf{ k})   =          2  \sum_{ 
              \mathbf{ p}    }^{    | \mathbf{ p}| < \Lambda  }   \frac{    2 E_{\mathbf{ p}}    L^3  \widetilde{G}^{(\mathbf{ P})}_{3\pi} ( \mathbf{ p} ,  \mathbf{ k} ;E)   }{  \frac{1}{V_R (\mu)}  - \widetilde{S}^{(\mathbf{ P})}_{3\pi} (   \mathbf{ k} ;E, \mu)  }  t^{(\mathbf{ P})}_{3\pi}(  \mathbf{ p})    , \label{3bLSeqdeltapot}
 \end{align}  
 where
 \begin{equation}
\widetilde{S}^{(\mathbf{ P})}_{3\pi} (   \mathbf{ k} ;E, \mu )   = \sum_{\mathbf{ p}  }^{    | \mathbf{ p}| < \Lambda  }  \left [   2 E_{\mathbf{ k} }    L^3  \widetilde{G}^{(\mathbf{ P})}_{3\pi} ( \mathbf{ p} ,  \mathbf{ k} ;E) -     \widetilde{G}^{(\mathbf{ 0})}_{2\pi} ( \mathbf{ p} ;\mu)  \right ]    .
\end{equation} 
The quantization condition of $3\pi^+$   can be rewritten with the renormalized coupling as
 \begin{align}
 & \det \left  [  \delta_{ \mathbf{ k}, \mathbf{ p}}   -   2    \frac{    2 E_{\mathbf{ p}}    L^3  \widetilde{G}^{(\mathbf{ P})}_{3\pi} ( \mathbf{ p} ,  \mathbf{ k} ;E)   }{  \frac{1}{V_R (\mu)}  - \widetilde{S}^{(\mathbf{ P})}_{3\pi} (   \mathbf{ k} ;E, \mu)  }    \right ] =0  ,  \nonumber \\
&\quad \quad  \quad \quad \quad \quad \quad   \quad \quad     (\mathbf{ k} ,\mathbf{ p} ) \in \frac{2\pi \mathbf{ n}}{L} , \mathbf{ n} \in \mathbb{Z}^3 ,\label{3bqcdeltapot}
 \end{align}  
which yields entire discrete energy spectrum of three pions in a finite  box  at continuum limit.

\section{Multi-$\pi^+$ dynamics with finite lattice spacing}\label{3bdynamicsfinitea}    

The lattice QCD simulations are usually performed in a cubic box with finite lattice spacing.  We discuss in this section impacts of finite lattice spacing and cubic lattice symmetry on multi-$\pi^+$ energy spectra.  We will subject the pion momenta to restriction imposed by the lattice spacing, but will not keep track of any other particular lattice artifacts, such as lattice action . The result will serve as a phenomenology motivated estimation of finite spacing effects on observables, especially the excited states, as we will see.

\subsection{Finite lattice spacing effect}\label{finitespacing}

With a finite lattice spacing $a$,  the coordinate of particles   become discrete,      a continuous integration   over coordinates  must be replaced by a discrete sum  over lattice sites:
\begin{equation}
    \int_{L^3} d \mathbf{ r} \to a^3 \sum_{\mathbf{ n}  } \, ,
\end{equation}
where the sum of $\mathbf{ n}$ is finite and is bound by the lattice spacing $a$ :
$$
(n_{x}, n_{y},n_{z}) \in [- N ,  N ] \; \text{and} \; N = \frac{L}{2 a} -1 \, .
$$
The infinite momentum sum  $\sum_\mathbf{p}$ where $\mathbf{ p} = \frac{2\pi }{L} \mathbf{ n}$ and $\mathbf{ n} \in  \mathbb{Z}^3$ at  continuum limit  is  replaced by a finite sum with momenta restricted  to the first Brillouin zone: $(n_{x}, n_{y},n_{z}) \in [- N ,  N ]$.

\begin{figure}
\begin{center}
\includegraphics[width=0.45\textwidth]{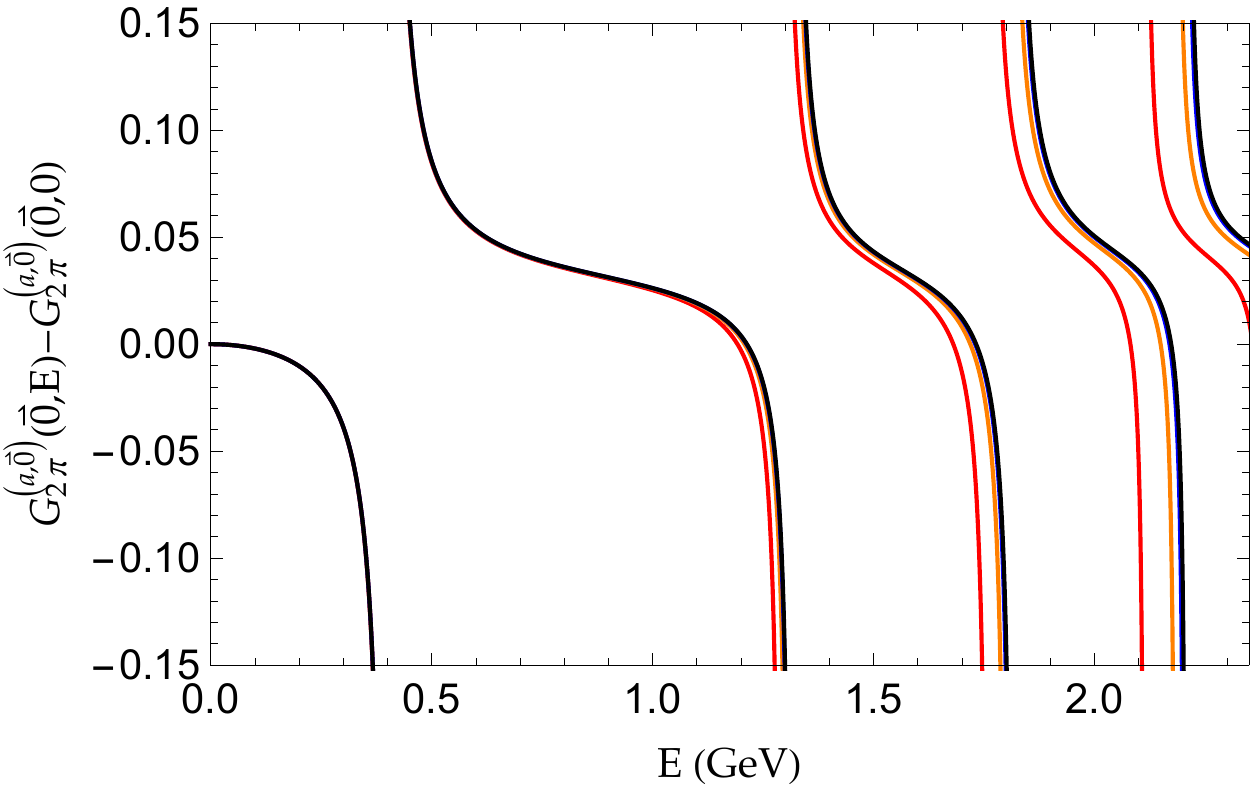}
\caption{ Plot of     $\sum_{\mathbf{ p}  }    \left [    \widetilde{G}^{(a, \mathbf{ 0})}_{2\pi} ( \mathbf{ p} ;E)  -    \widetilde{G}^{(a, \mathbf{ 0})}_{2\pi} ( \mathbf{ p} ;\mu) \right ]$   for various finite lattice spacing $a$'s  with  $\mu= 0 \mbox{GeV}$ $L=10 \mbox{GeV}^{-1}$ and $m_\pi=0.200 \mbox{GeV}$: $a= 0.48\mbox{GeV}^{-1}$ (Red), $0.24\mbox{GeV}^{-1}$ (Orange), $0.12\mbox{GeV}^{-1}$ (Blue), and $ 0\mbox{GeV}^{-1}$ (Black). }\label{G2bfinitespacplot}
\end{center}
\end{figure}

\begin{figure}
\begin{center}
\includegraphics[width=0.45\textwidth]{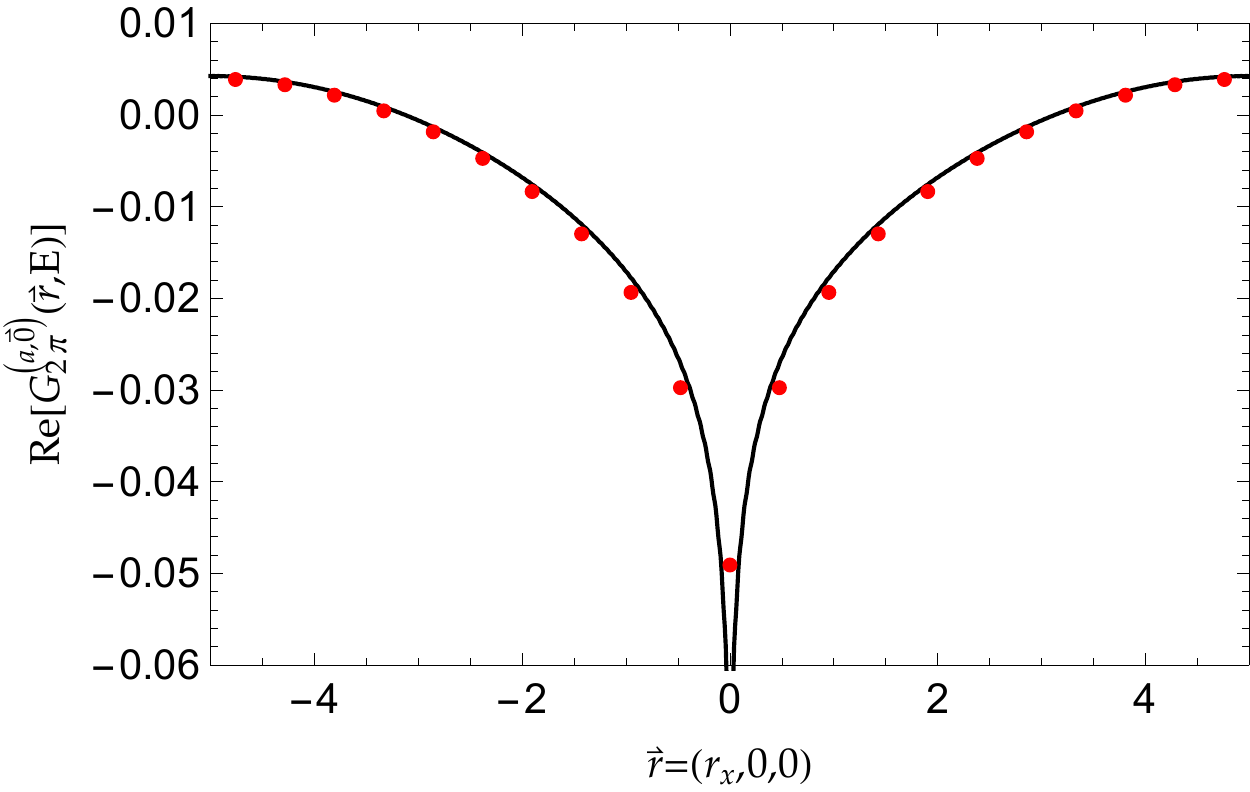}
\caption{ Plot of real part of  finite spacing version of   $  G^{(a, \mathbf{ 0})}_{2\pi} ( \mathbf{ r}  ;E)  = \sum_{\mathbf{ p}  }  e^{i \mathbf{ p} \cdot \mathbf{ r}}  \widetilde{G}^{(a, \mathbf{ 0})}_{2\pi} ( \mathbf{ p} ;E)   $ (red dots)  compared with its infinite volume counterpart  (black curve)  with chosen parameters: $\mathbf{ r} = (r_x,0,0)$,  $E= 1 \mbox{GeV}$ $L=10 \mbox{GeV}^{-1}$, $m_\pi=0.200 \mbox{GeV}$,   and $a =0.48\mbox{GeV}^{-1}$. }\label{G2brxplot}
\end{center}
\end{figure}

In addition, the continuous relativistic energy momentum dispersion relation, $E_\mathbf{ p} = \sqrt{m_\pi^2 + \mathbf{ p}^2}$, is replaced by lattice dispersion relation with a explicit dependence on the finite lattice spacing $a$,
\begin{equation}
2 \sinh \frac{a E_\mathbf{ p}}{2} =\sqrt{ 2 \cosh a m_\pi + 4 - 2 \sum_{i=x,y,z} \cos a p_i }, 
\end{equation}
 where $p_i  = \frac{2 \pi}{L} n_i$, $i=x,y,z$  and $n_i \in [-N,N]$. Therefore, to convert all  the continuum limit multi-$\pi^+$ dynamical equations presented in Section \ref{3bdynamics} to equations with explicit finite lattice spacing effect build in, the following relations between  the physical quantities at continuum limit  and  their finite lattice spacing counterparts   must be considered: 
 \begin{equation}
 E  \leftrightarrow \frac{2}{a} \sinh \frac{a E }{2},  \   m_\pi \leftrightarrow  \frac{2}{a} \sinh \frac{a  m_\pi}{2}, \  p_{i} \leftrightarrow \frac{2}{a} \sin \frac{a p_{i}}{2}.
 \end{equation}
  For examples, the momentum space Green's function defined in Eq.(\ref{g2bcont}) and Eq.(\ref{g3bcont}) are now replaced by finite lattice spacing correspondences,
 \begin{align}
&  \widetilde{G}^{(a,\mathbf{ P})}_{2\pi} ( \mathbf{ p} ;E)   = \frac{a^3}{L^3}  \frac{2 (2 \sinh \frac{a E_\mathbf{ p} }{2} +2 \sinh \frac{a E_{\mathbf{ P}-\mathbf{ p}}}{2})}{4 \sinh \frac{a  E_\mathbf{ p} }{2} 4 \sinh \frac{a E_{\mathbf{ P}-\mathbf{ p} } }{2} } \nonumber \\
 & \times \frac{1}{ (2 \sinh \frac{aE}{2})^2- (2 \sinh \frac{a E_\mathbf{ p} }{2} +2 \sinh \frac{a E_{\mathbf{ P}-\mathbf{ p}}}{2})^2 }, \label{g2bfv}
 \end{align}
for  $2\pi^+$, and   
  \begin{align}
 \widetilde{G}^{(a,\mathbf{ P})}_{3\pi} & ( \mathbf{ p}_1 ,  \mathbf{ p}_2 ;E)    = \frac{a^4}{L^6}  \frac{2(  \sum_{i=1}^3 2 \sinh \frac{a E_{\mathbf{ p}_i} }{2}  )}{4 \sinh \frac{a E_{\mathbf{ p}_1} }{2}  4 \sinh \frac{a E_{\mathbf{ p}_2} }{2}  4 \sinh \frac{a E_{ \mathbf{ p}_3} }{2} } \nonumber \\
& \times \frac{1}{(2 \sinh \frac{aE}{2})^2-  ( \sum_{i=1}^3 2 \sinh \frac{a E_{\mathbf{ p}_i} }{2}  )^2 },\label{g3bfv}
 \end{align}
for $3\pi^+$, where $\mathbf{ p}_3 =\mathbf{ P}-\mathbf{ p}_1 -\mathbf{ p}_2 $. The symbol $a$ in superscript is used to label  the functions defined with a finite lattice spacing.  The examples of finite lattice spacing effects  are illustrated in Fig.~\ref{G2bfinitespacplot} and Fig.~\ref{G2brxplot}.

\subsubsection{$2\pi^+$ LS equation with finite lattice spacing}
With all the ingredients  mentioned previously,   the $2\pi^+$ LS equation for a general potential  at a finite lattice spacing is obtained by replacing  Eq.(\ref{2bLSeq})  by
 \begin{align}
 t^{(a,\mathbf{ P})}_{2\pi}  (\mathbf{ k} )  &=    \sum_{\mathbf{ p}  }    \widetilde{V}( | \mathbf{ k} - \mathbf{ p} |)   \widetilde{G}^{(a, \mathbf{ P})}_{2\pi}  ( \mathbf{ p} ;E) t^{(a,\mathbf{ P})}_{2\pi}  (\mathbf{ p}) ,  \nonumber \\
 & \quad    ( \mathbf{ k}, \mathbf{ p}) \in \frac{2\pi \mathbf{ n}}{L}, \  ( n_{x },n_{y },n_{z } ) \in [-N,N]. \label{2bLSeqfinitea}
 \end{align}
For contact interaction,  a finite lattice spacing may play the role of natural ultraviolet cutoff,  Eq.(\ref{2bqcrenorm}) is thus replaced by   
 \begin{equation}
 \frac{1}{V_R (\mu)}  =    \sum_{\mathbf{ p}  }      \widetilde{G}^{(a,\mathbf{ P})}_{2\pi} ( \mathbf{ p} ;E)  - \sum_{\mathbf{ p}  }      \widetilde{G}^{(a,\mathbf{ 0})}_{2\pi} ( \mathbf{ p} ;\mu) , \label{2bqcfinitea}
 \end{equation}
where sum of $\mathbf{ p} = \frac{2\pi \mathbf{ n}}{L} $ is restricted in first  Brillouin zone with $(n_{x}, n_{y},n_{z}) \in [- N ,  N ]$ and $N= \frac{L}{2a}-1$.

\subsubsection{$3\pi^+$ LS equation with finite lattice spacing}
With same strategy,   the finite lattice spacing version of $3\pi^+$ LS equation is obtained by replacing   Eq.(\ref{3bLSeqdeltapot})   by
  \begin{align}
 t^{(a,\mathbf{ P})}_{3\pi}( \mathbf{ k}) &   =          2  \sum_{ 
              \mathbf{ p}    }    \frac{    4  \sinh \frac{a E_\mathbf{ p}}{2}  \frac{L^3}{a}  \widetilde{G}^{(a, \mathbf{ P})}_{3\pi} ( \mathbf{ p} ,  \mathbf{ k} ;E)   }{  \frac{1}{V_R (\mu)}  - \widetilde{S}^{(a, \mathbf{ P})}_{3\pi} (   \mathbf{ k} ;E, \mu)  }  t^{(a,\mathbf{ P})}_{3\pi}(  \mathbf{ p})    , \nonumber \\
 & \quad \quad  ( \mathbf{ k}, \mathbf{ p}) \in \frac{2\pi \mathbf{ n}}{L}, \ (n_{x}, n_{y},n_{z})  \in [-N,N], \label{3bLSeqfinitea}
 \end{align}  
 where
\begin{align}
&\widetilde{S}^{(a, \mathbf{ P})}_{3\pi} (   \mathbf{ k} ;E, \mu )  \nonumber \\
  & =   \sum_{\mathbf{ p}    }  \left [   4  \sinh \frac{a E_\mathbf{ k}}{2}   \frac{ L^3 }{a}  \widetilde{G}^{(a, \mathbf{ P})}_{3\pi} ( \mathbf{ p} ,  \mathbf{ k} ;E) -     \widetilde{G}^{(a, \mathbf{ 0})}_{2\pi} ( \mathbf{ p} ;\mu)  \right ]   ,
\end{align}
 and   $\mathbf{ p} = \frac{2\pi \mathbf{ n}}{L}, \ (n_{x}, n_{y},n_{z})  \in [-N,N]$.

\subsection{Cubic lattice symmetry group and its irreducible representations }\label{irrepproj} 

The energy spectrum of a quantum system  is normally organized and labeled  according to irreducible representations (irreps) of the symmetry  groups of the system.  These irreps carry so-called ``good'' quantum numbers that help in practice identify states of the system. For example, in infinite volume, hadronic bound states are typically labeled in terms of total angular momentum and parity based on   the system's behavior under rotations and space inversion. To  decouple states with different quantum numbers, one can project the dynamic equations of the system onto each  irrep of symmetry  groups.
The end result is that each eigen-energy   belongs to certain irrep, or irreps in the case of degeneracy.

Similar operations  can be carried out  for finite-volume systems as well.
The energy  spectra of finite-volume multi-$\pi^+$ system  are expected to be labeled by irreps of the  cubic lattice symmetry group. For  instance, the cubic symmetry group for  a system with vanishing total momentum, $\mathbf{ P}=0$, is the octahedral group $O_h$, which consists of 48 symmetry operations, including 24 discrete space rotations and  inversions of all axes. The irreps of octahedral group $O_h$  include   one-dimensional representations, $A^{\pm}_1$ and $A^{\pm}_2$,   two-dimensional representation, $E^{\pm}$, and   three-dimensional representations, $T^{\pm}_1$ and $T^{\pm}_2$. The superscripts $\pm$ are used to label even or odd parity state of the system.  A brief introduction on the subject of  irrep projection of dynamical  equations is given in  this section,   with the $2\pi^+$ LS equation   used as a specific example.  More elaborate explanations on the subject  can be found in, e.g., Refs.~\cite{Doring:2018xxx,Cornwell:1997ke}. The  irrep-projected $2\pi^+$ and $3\pi^+$ LS equations in  the CM frame will be presented in   the following subsections.

 We start with the projection operator for the cubic group  \cite{Cornwell:1997ke}, 
 \begin{equation}
 \mathcal{P}^{(\lambda)}_{\alpha,\alpha} = \frac{d_\lambda}{48} \sum_{g \in \mathcal{G}} \Gamma_{\alpha,\alpha}^{(\lambda) *} (g)  \mathbf{ O}(  g), \label{projops}
 \end{equation}
where  $\mathcal{G}$ and $g$ stand for the  the cubic symmetry group and its  elements.  Here $\lambda$ is used to label a specific irrep, and  $d_\lambda$ denotes the dimension of irrep $\lambda$, {\it e.g.},  $d_\lambda = 1,1,2,3,3$ for  $\lambda= A^{\pm}_1, A^{\pm}_2, E^{\pm}, T^{\pm}_1,T^{\pm}_2$, respectively.   With $g$ running over all elements of $\mathcal{G}$, $\Gamma^{(\lambda) } (g) $ 
are a set of $d_\lambda$-by-$d_\lambda$ matrices that furnish irrep $\lambda$.  $ \mathbf{ O}(  g)$ represents the symmetry operation implemented on  quantum states. These operations are perhaps most easily explained by their action on momenta or coordinates of the particles. For instance, 
$$
  \mathbf{ p}'= \mathbf{ O}(  g) \mathbf{ p} =  g \mathbf{ p} \, ,
$$
where $|\mathbf{ p}' | = |\mathbf{ p}|$ for all $g \in \mathcal{G}$, 
because operations in $O_h$ are either rotation or inversion. The complete list of symmetry operations of the octahedral group $O_h$  are given in  Appendix~\ref{chartable}.

Momenta   of a particle can be grouped  into various sets, and momenta in the same set are connected by symmetry operations. Any set of such momenta can be represented by a single reference 
vector $\mathbf{ p}_0$,  and the rest of members of this particular set can be reached by 
$$
\mathbf{ p} = g \mathbf{ p}_0\quad  (g\in \mathcal{G}) \, .
$$
The reference vector used in this work is   the same concept as used in Ref.~\cite{Doring:2018xxx},   while in Ref.~\cite{Cornwell:1997ke} the set of momenta    represented by $\mathbf{ p}_0$ is called the ``star'' of $\mathbf{ p}_0$.  
With a  cutoff on the lattice momenta, 
$$
\mathbf{ p} = \frac{2\pi \mathbf{ n}}{L} \, ,
\quad (n_{x}, n_{y},n_{z})  \in [-N,N] \, ,
$$
the reference vectors, $\mathbf{ p}_0$'s, may be  chosen as $$
\mathbf{ p}_0 \in \{\frac{2\pi}{L}  \mathbf{ n}_{i,j,k} \},
$$ 
where  
$$
\mathbf{ n}_{i,j,k}   = (k, j, i) \, ,
$$
and   $(i,j,k) \in [0,N]$ satisfies $k \leq j \leq i$.  The total number of reference vectors $\mathbf{ p}_0$ for a fixed $N$ is  thus given by   
$$
\sum_{i=0}^N \sum_{j=0}^i \sum_{k=0}^j = \frac{1}{6}(N+1)(N+2)(N+3) \, .
$$
In terms of   reference vectors and symmetry  operations, the sum of momenta over an arbitrary function can be reorganized  as follows:
  \begin{equation}
  \sum_{\mathbf{ p}  }  f (\mathbf{ p})   = \sum_{\mathbf{ p}_0}  \frac{\vartheta(\mathbf{ p}_0)}{48}  \sum_{g  \in \mathcal{G}} f (g \mathbf{ p}_0), \label{multiplicity}
  \end{equation}
  where  $\vartheta(\mathbf{ p}_0)$  is the multiplicity of distinct   momenta within the set  represented by $\mathbf{ p}_0$.  For instance, for
  $$
  \mathbf{ p}_0 = \frac{2\pi}{L} (0,0,1)\, , 
  $$
  $\vartheta(\mathbf{ p}_0) =6 $
  and   the six distinct momenta are
  $$
  \mathbf{ p}= g\mathbf{ p}_0   \in \frac{2\pi}{L} \{ (0,0, \pm 1), \ (0, \pm 1,0), \ ( \pm 1,0,0) \}\, .
 $$
The multiplicity function $\vartheta(\mathbf{ p}_0)$ used in this work has the same meaning as the multiplicity of a given shell defined in Ref.~\cite{Doring:2018xxx}. 
 Therefore, the projector defined in Eq.(\ref{projops})  acting on  amplitudes can be interpreted as weighted  average within  the momentum set represented by $\mathbf{ p}_0$, and the projected amplitudes  can be labeled   by   $\mathbf{ p}_0$'s.

 \subsection{Irrep projection of $2\pi^+$ LS equation }\label{2pionirrepproj} 

As  an example,  we discuss projection of two-pion amplitudes for $\mathbf{ P}=\mathbf{0}$ onto irrep $\lambda$  that is  given by
\begin{equation}
 t^{( \lambda )}_{2\pi}   (\mathbf{ k}_0 )  = \mathcal{P}^{(\lambda)}_{\alpha,\alpha}  t^{(\mathbf{ 0})}_{2\pi}  (\mathbf{ k} )  =  \frac{d_\lambda}{48} \sum_{g \in \mathcal{G}} \Gamma_{\alpha,\alpha}^{(\lambda) *} (g)  t^{(\mathbf{ 0})}_{2\pi}  (g \mathbf{ k}_0 ) \, .
\end{equation}
 The projected amplitude, $t^{( \lambda )}_{2\pi}   (\mathbf{ k}_0 )  $,   actually does not depend on quantum number $\alpha$ due to  the cubic symmetry of the system. Therefore,
 it may be convenient to  express the projection operator in terms of the character of irreps:
\begin{equation}
\mathcal{P}^{(\lambda)} = \frac{1}{d_\lambda} \sum_{\alpha} \mathcal{P}^{(\lambda)}_{\alpha,\alpha}   = \frac{1}{48} \sum_{g \in \mathcal{G}} \chi^{(\lambda) *} (g)    \mathbf{ O}(  g),
\end{equation}
 where 
 $$
 \chi^{(\lambda) } (g) = \sum_{\alpha}  \Gamma_{\alpha,\alpha}^{(\lambda) } (g) 
 $$
 is the character of irrep $\lambda$, and   it satisfies orthogonality relation:
\begin{equation}
\frac{1}{48} \sum_{g \in \mathcal{G}} \chi^{(\lambda) *} (g)   \chi^{(\lambda') } (g) = \delta_{\lambda, \lambda'} .  
\end{equation}
 With the help of $\mathcal{P}^{(\lambda)}$,  one can rewrite the projection of two-pion amplitudes in a more compact form:
 \begin{equation}
 t^{( \lambda )}_{2\pi}  (\mathbf{ k}_0)  = \frac{1}{48} \sum_{g \in \mathcal{G}} \chi^{(\lambda) *} (g)     t^{(\mathbf{ 0})}_{2\pi}  ( g \mathbf{ k}_0) . \label{projopschar}
\end{equation}
The above equation can be inverted using the orthogonality relation:
 \begin{equation}
     t^{(\mathbf{ 0})}_{2\pi}  (  g \mathbf{ k}_0) = \sum_\lambda  \chi^{(\lambda) } (g)   t^{( \lambda )}_{2\pi}   (\mathbf{ k}_0)  . \label{projexpansion}
\end{equation}

Projection of dynamic equations established in previous sections is obtained by applying projection formula of amplitudes.
Applying Eqs.~\eqref{projopschar}, \eqref{projexpansion}, and \eqref{multiplicity} to Eq.~\eqref{2bLSeq}, we arrive at
 \begin{equation}
   t^{( \lambda )}_{2\pi}   (\mathbf{ k}_0) =    \sum_{\mathbf{ p}_0  } \vartheta(\mathbf{ p}_0)   \widetilde{V}^{(\lambda)} (\mathbf{ k}_0, \mathbf{ p}_0 )     \widetilde{G}^{(\mathbf{ 0})}_{2\pi}  ( \mathbf{ p}_0 ;E)   t^{( \lambda)}_{2\pi}   (\mathbf{ p}_0)    , \label{2bLSirrep}
 \end{equation}
 where   $ \widetilde{V}^{(\lambda)} (\mathbf{ k}_0, \mathbf{ p}_0 )  $ is  the projected potential:
 \begin{align}
&  \delta_{\lambda, \lambda'} \widetilde{V}^{(\lambda)} (\mathbf{ k}_0, \mathbf{ p}_0 )  \nonumber \\
& = \frac{1}{48^2}  \sum_{g_\mathbf{ p} , g_\mathbf{ k}  \in \mathcal{G}}     \chi^{(\lambda) *} (g_\mathbf{ k})     \widetilde{V}( | g_\mathbf{ k} \mathbf{ k}_0 - g_\mathbf{ p} \mathbf{ p}_0 |  )  \chi^{(\lambda') } (g_\mathbf{ p}  )  \nonumber \\
& =\delta_{\lambda, \lambda'}  \frac{1}{48}  \sum_{  g  \in \mathcal{G}}  \chi^{(\lambda) *} ( g)       \widetilde{V}( |  g  \mathbf{ k}_0 -  \mathbf{ p}_0 | )  .
 \end{align}
The energy spectrum stemming from the irrep-projected dynamical equation is of course labeled by the same irrep.

With the contact interaction,  $\widetilde{V} (k)= V_0$,     only   $A^+_1$ irrep survives after projection of  the potential:
$$
\widetilde{V}^{(\lambda)} (\mathbf{ k}_0, \mathbf{ p}_0 ) = \delta_{\lambda, A_1^+} V_0 \, .
$$
Therefore, projection of Eq.(\ref{2bqcfinitea})  yields non-trivial solutions only in $A_1^+$:  
  \begin{equation}
 \frac{\delta_{\lambda, A_1^+} }{V_R (\mu)}  =    \sum_{\mathbf{ p}_0  }    \vartheta(\mathbf{ p}_0)   \widetilde{G}^{(a,\mathbf{ 0})}_{2\pi}  ( \mathbf{ p}_0 ;E)  - \sum_{\mathbf{ p}_0  } \vartheta(\mathbf{ p}_0)      \widetilde{G}^{(a, \mathbf{ 0})}_{2\pi}  ( \mathbf{ p}_0 ;\mu) .  \label{2bqcirrep}
 \end{equation}

\subsection{Irrep projection of $3\pi^+$ LS equation }\label{3pionirrepproj}  

 Since the three-pion amplitudes depend on two  momentum variables,  the   irrep projection   can be done by first projecting out each momentum dependence  separately onto  the corresponding irrep,   and then coupling two individual projections  to  an irrep of the $3\pi^+$ system.  
 This procedure resembles  addition of angular momenta, which is essentially reduction of tensor product of two $SO(3)$ irreps.

Following that idea,  projection of $3\pi^+$  amplitude $t^{(\mathbf{ 0})}_{(k)} ( \mathbf{ k}_1, \mathbf{ k}_2)$  is accomplished by  the following operator:
\begin{align}
\mathcal{P}^{(\lambda)}_{\alpha} (1,2) = \sum_{\alpha_1, \alpha_2}  \left (  \begin{array}{cc}  \lambda_1 & \lambda_2 \\ \alpha_1 & \alpha_2 \end{array}  \left |   \begin{array}{cc}  \lambda  \\ \alpha    \end{array}   \right.    \right ) \mathcal{P}^{(\lambda_1)}_{\alpha_1, \alpha_1} (1)   \mathcal{P}^{(\lambda_2)}_{\alpha_2, \alpha_2} (2),
\end{align}
where $\mathcal{P}^{(\lambda_1)}_{\alpha_1, \alpha_1} (1)  $ and $ \mathcal{P}^{(\lambda_2)}_{\alpha_2, \alpha_2} (2)$ are the  uncoupled irrep projection operators on momenta $ \mathbf{ k}_1 $ and $ \mathbf{ k}_2$, respectively, and  
$$
\left (  \begin{array}{cc}  \lambda_1 & \lambda_2 \\ \alpha_1 & \alpha_2 \end{array}  \left |   \begin{array}{cc}  \lambda  \\ \alpha    \end{array}   \right.    \right ) 
$$
is the Clebsch-Gordan coefficient  that couple irreps $\lambda_1$ and $\lambda_2$ to   irrep $\lambda$ on  the $\alpha$-th row.  Again,  due to the   cubic symmetry of the system, projected three-body amplitudes  do not depend on   $\alpha$,   so the projection of  $t^{(\mathbf{ 0})}_{(k)} ( \mathbf{ k}_1, \mathbf{ k}_2)$   can be written as
\begin{align}
&   t^{(\lambda)}_{(k)} ( \mathbf{ k}_{10}, \mathbf{ k}_{20})   =  \frac{1}{d_\lambda} \sum_{\alpha}  \mathcal{P}^{(\lambda)}_{\alpha} (1,2)  t^{(\mathbf{ 0})}_{(k)}  ( \mathbf{ k}_1, \mathbf{ k}_2) \nonumber \\
& = \frac{d_{\lambda_1} d_{\lambda_2} }{48^2}    \frac{1}{d_\lambda}   \sum_{\alpha_1, \alpha_2, \alpha }  \left (  \begin{array}{cc}  \lambda_1 & \lambda_2 \\ \alpha_1 & \alpha_2 \end{array}  \left |   \begin{array}{cc}  \lambda  \\ \alpha    \end{array}   \right.    \right )        \nonumber \\
&         \times \sum_{g_1, g_2 \in \mathcal{G}}   \Gamma_{\alpha_1,\alpha_1}^{(\lambda_1) *} (g_1)      \Gamma_{\alpha_2,\alpha_2}^{(\lambda_2) *} (g_2)  t^{(\mathbf{ 0})}_{(k)}  ( g_1 \mathbf{ k}_{10}, g_2 \mathbf{ k}_{20})  .
\end{align}

The  general projection of three-particle amplitudes can be cumbersome. However for the  interaction under consideration in the present paper, no three-body force and only pair-wise contact interaction, the procedure is greatly simplified.
 This is because, as shown in Sec. \ref{contactpot},  
 the three-pion amplitude depends only on a single momentum:  $t^{(\mathbf{ P})}_{3\pi} ( \mathbf{ k}_2)$.   With no dependence on $\mathbf{ k}_1$, only  the trivial irrep $A_1^+$ survives after projection  of $\mathbf{ k}_1$ dependence:   $\lambda_1= A_1^+$  and $\Gamma_{\alpha_1,\alpha_1}^{(\lambda_1) }  =1$.     The  coupled irrep of the $3\pi^+$ system is thus determined by $\mathbf{ k}_2$ dependence  alone:  $\lambda = \lambda_2$ and $\alpha =\alpha_2$.  Therefore, one can apply straightforwardly most of the results for the two-pion case in Sec.~\ref{2pionirrepproj} to the irrep projection of the $3\pi^+$ LS equation.

With pair-wise contact interaction,   the  irrep projection of   $t^{(\mathbf{ 0})}_{3\pi} ( \mathbf{ k}_2)$  in the CM frame is given by  
 \begin{equation}
 t^{( \lambda )}_{3\pi}  (\mathbf{ k}_0)  = \frac{1}{48} \sum_{g \in \mathcal{G}} \chi^{(\lambda) *} (g)     t^{(a,\mathbf{ 0})}_{3\pi}  ( g \mathbf{ k}_0) \, .
\end{equation}
 Similar to  Eq.(\ref{projexpansion}), we also have
 \begin{equation}
     t^{(a,\mathbf{ 0})}_{3\pi}  (  g \mathbf{ k}_0) = \sum_\lambda  \chi^{(\lambda) } (g)   t^{( \lambda )}_{3\pi}   (\mathbf{ k}_0)  .
\end{equation}
Noticing that $\widetilde{S}^{(a, \mathbf{ 0})}_{3\pi} (   \mathbf{ k}_0 ;E, \mu) $  remains invariant under symmetry  operations,
 \begin{equation}
 \widetilde{S}^{(a, \mathbf{ 0})}_{3\pi} (   g \mathbf{ k}_0 ;E, \mu)   = \widetilde{S}^{(a, \mathbf{ 0})}_{3\pi} (   \mathbf{ k}_0 ;E, \mu)  ,
 \end{equation}
 and   that the projection on the  right-hand side of Eq.(\ref{3bLSeqfinitea}) yields 
  \begin{align}
&      \frac{1}{48^2} \sum_{g_1, g_2  \in \mathcal{G}} \chi^{(\lambda) *} (g_1)    \chi^{(\lambda') } (g_2)     \nonumber \\
&       \times \left [ 4  \sinh \frac{a E_{g_2 \mathbf{ p}_0 }}{2}   \frac{L^3}{a}  \widetilde{G}^{(a, \mathbf{ 0})}_{3\pi} ( g_2 \mathbf{ p}_0 , g_1  \mathbf{ k}_0 ;E)  \right ]  \nonumber \\
&=   \delta_{\lambda,\lambda'}  \frac{1}{48} \sum_{g   \in \mathcal{G}} \chi^{(\lambda) *} (g )      4  \sinh \frac{a E_{  \mathbf{ p}_0 }}{2}       \frac{L^3}{a}  \widetilde{G}^{(a, \mathbf{ 0})}_{3\pi} (   \mathbf{ p}_0 , g  \mathbf{ k}_0 ;E)         ,
 \end{align}
 one can show that irrep projection of the $3\pi^+$ LS equation ~\eqref{3bLSeqfinitea} ultimately leads to  the following:
 \begin{align}
   t^{( \lambda )}_{3\pi}  (\mathbf{ k}_0)     
=    2  \sum_{ 
              \mathbf{ p}_0    }  \frac{  \vartheta(  \mathbf{ p}_0)   \widetilde{C}^{(\lambda)}_{3\pi} (   \mathbf{ k}_0 ,  \mathbf{ p}_0 ;E) }{ \frac{1}{V_R (\mu)}  - \widetilde{S}^{(a, \mathbf{ 0})}_{3\pi} (     \mathbf{ k}_0 ;E, \mu)  } t^{( \lambda )}_{3\pi}   ( \mathbf{ p}_0)     , \label{3bLSproj}
 \end{align} 
 where  
 \begin{align}
  & \widetilde{C}^{(\lambda)}_{3\pi} (   \mathbf{ k}_0 ,  \mathbf{ p}_0 ;E)   \nonumber \\
&   =   \frac{1}{48} \sum_{g   \in \mathcal{G}} \chi^{(\lambda) *} (g )        4  \sinh \frac{a E_{  \mathbf{ p}_0 }}{2}       \frac{L^3}{a}  \widetilde{G}^{(a, \mathbf{ 0})}_{3\pi} (   \mathbf{ p}_0 , g  \mathbf{ k}_0 ;E)           .
 \end{align}
 The $O_h$-irrep projected $3\pi^+$ quantization condition, under our assumptions,  has a simple form:
 \begin{equation}
\det \left [ \delta_{\mathbf{ k}_0,\mathbf{ p}_0 } - \frac{ 2  \vartheta(  \mathbf{ p}_0)   \widetilde{C}^{(\lambda)}_{3\pi} (   \mathbf{ k}_0 ,  \mathbf{ p}_0 ;E) }{\frac{1}{V_R (\mu)}  - \widetilde{S}^{(a, \mathbf{ 0})}_{3\pi} (     \mathbf{ k}_0 ;E, \mu)  } \right ]=0. \label{3bqcirrep}
\end{equation} 
The  above equation suggests  that except  for $A_1^-$ irrep, only trivial solutions   --- free particle states--- can be found near the ground state with all three pions at rest,  
because the only irrep in which $ \widetilde{C}^{(\lambda)}_{3\pi} (   \mathbf{ 0} ,  \mathbf{ 0} ;E) $ does not vanish is $\lambda = A_1^-$.

\begin{figure*}
\centering
\begin{subfigure}[b]{0.32\textwidth}
\includegraphics[width=\textwidth]{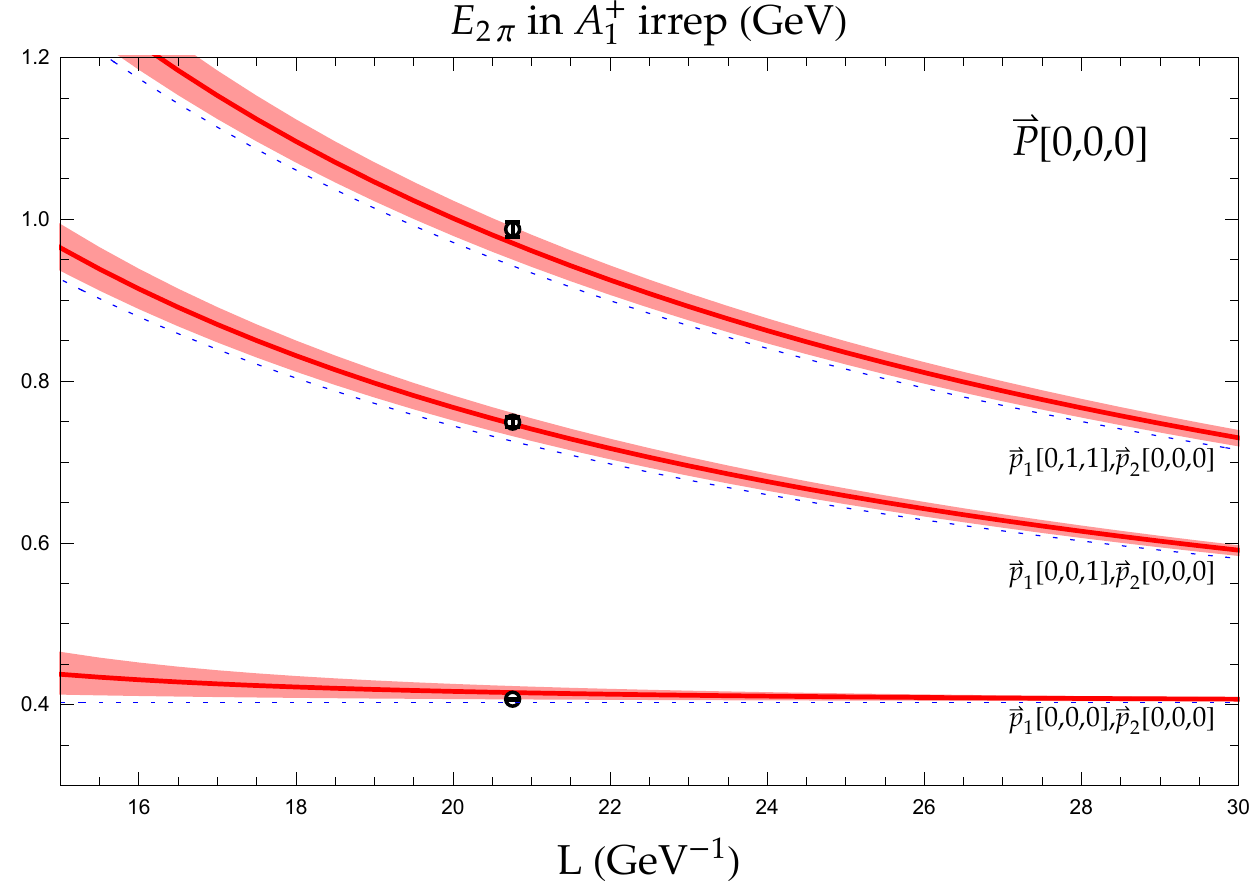}
\caption{ $2\pi^+$ energy spectrum   in $A^+_1$ irrep.}
\end{subfigure}
\begin{subfigure}[b]{0.32\textwidth}
\includegraphics[width=\textwidth]{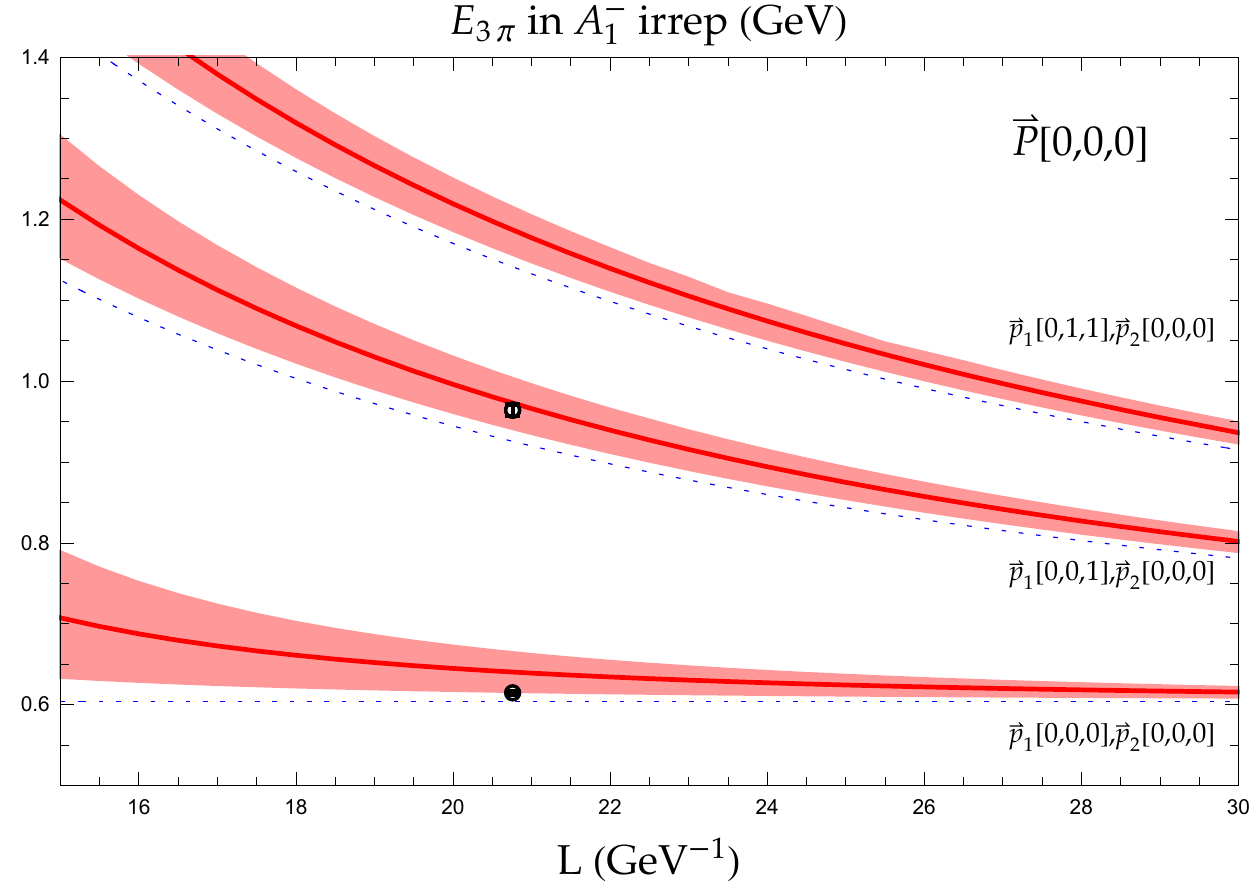}
\caption{ $3\pi^+$ energy spectrum   in $A^-_1$ irrep.}
\end{subfigure}
\begin{subfigure}[b]{0.32\textwidth}
\includegraphics[width=\textwidth]{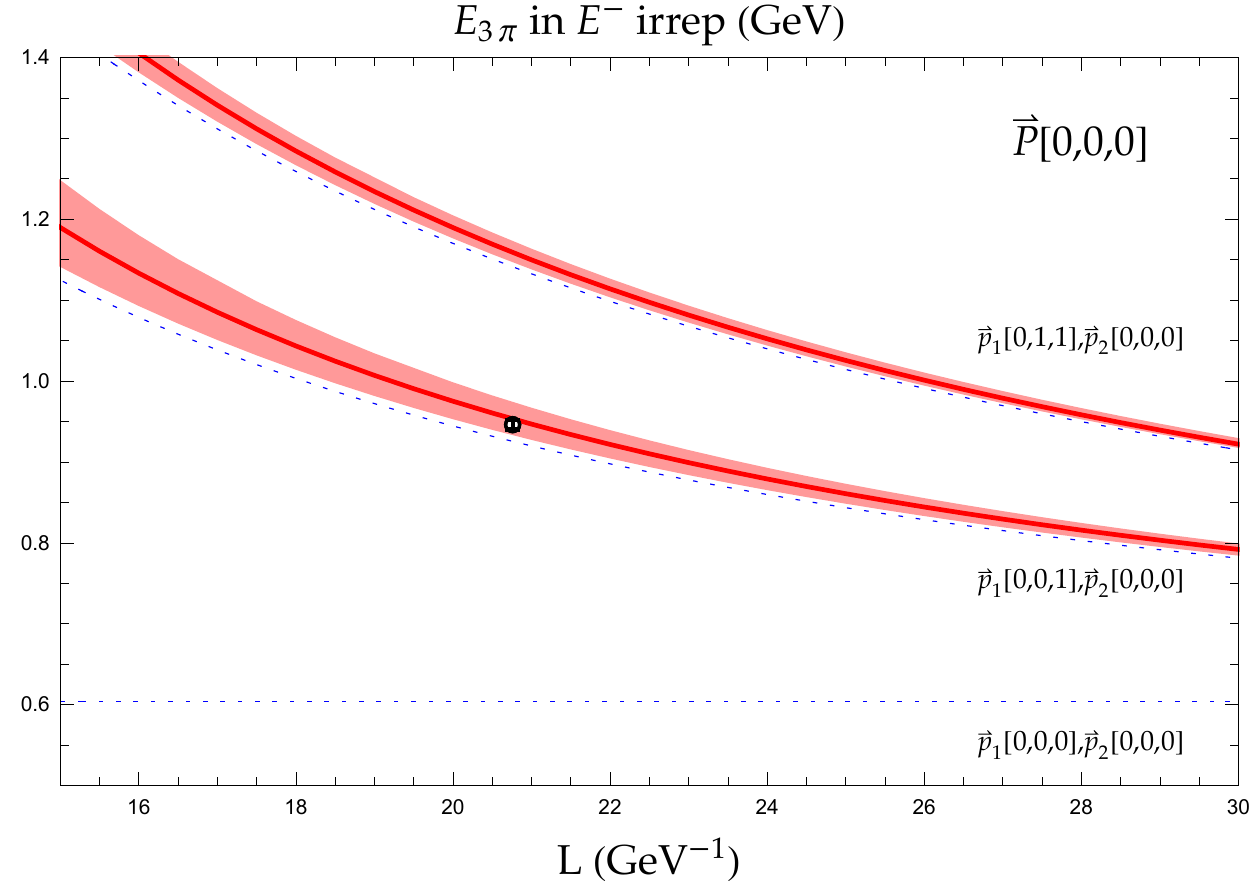}
\caption{$3\pi^+$ energy spectrum in $E^-$ irrep.}
\end{subfigure}
\caption{ $2\pi^+$ and $3\pi^+$  energy spectrum in irreps $A_1$ and $E$ and CM frame: red bands are spectrum produced by using  Eq.(\ref{2bqcirrep}) and  Eq.(\ref{3bqcirrep}) for $2\pi^+$ and $3\pi^+$ respectively, the lattice results (black  circles) are taken from \cite{Horz:2019rrn}.   The free multi-$\pi^+$ energy spectrum (blue dashed curves) by using $E=\frac{2}{a} \sinh^{-1} ( \sum_{i} \sinh \frac{a E_{\mathbf{ p}_i}}{2})$ and $\sum_i \mathbf{ p}_i = \mathbf{ P} $ are also plotted as reference.  The lattice spacing used in this work and in \cite{Horz:2019rrn} is  $a=0.324 \mbox{GeV}^{-1}$, $m_\pi = 0.200 \mbox{GeV}$.  }
\label{E2b3bplot}
\end{figure*}

\section{Numerical results}\label{numerics}
 
In this section, we use the quantization conditions \eqref{2bqcirrep} and \eqref{3bqcirrep} to produce $2\pi^+$ and $3\pi^+$ energy spectra for various lattice sizes. A single  parameter, the  renormalized coupling of the two-body contact interaction $V_R(\mu)$, is fitted to  the lattice results  of Ref.~\cite{Horz:2019rrn}. 

\begin{figure}
\begin{center}
\includegraphics[width=0.49\textwidth]{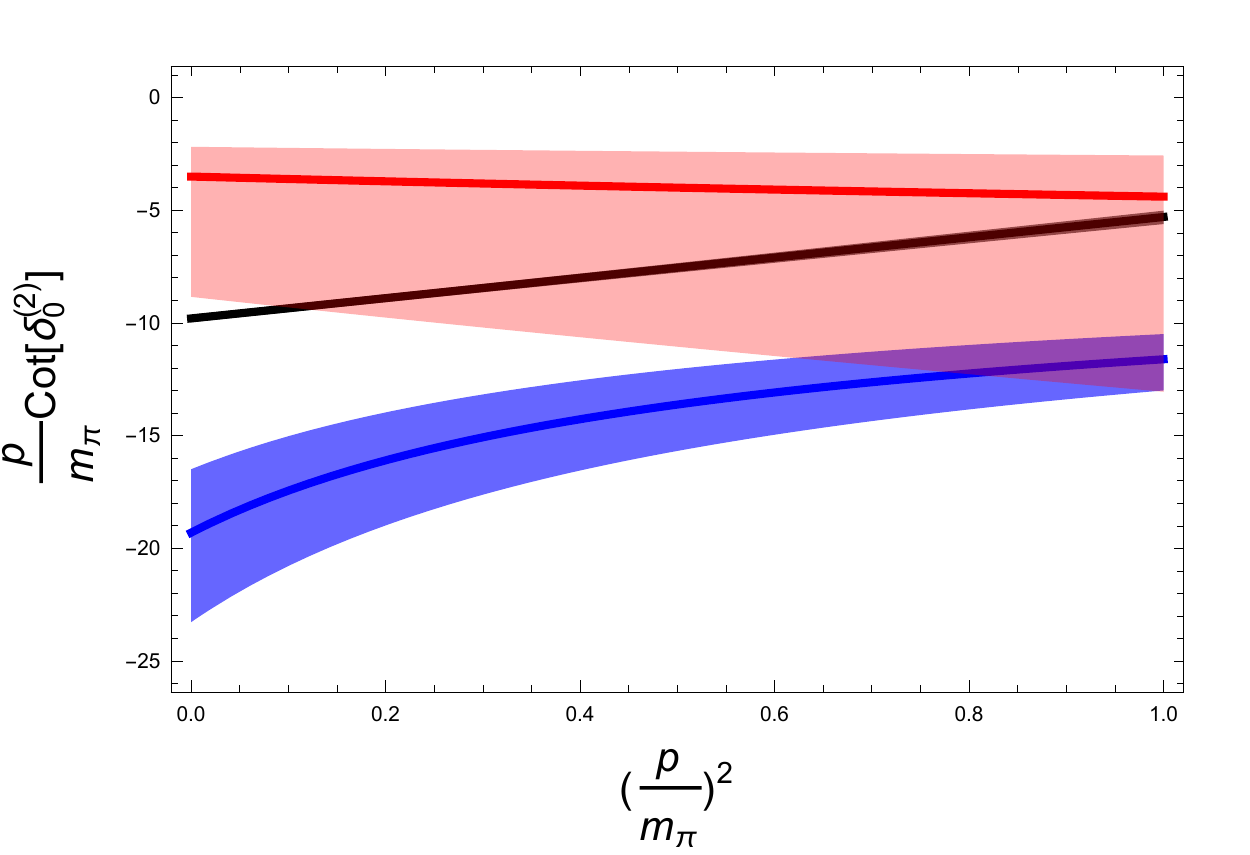}
\caption{ The plot of $\frac{p}{m_\pi}\cot \delta_0^{(2)}(E)$ vs. $(\frac{p}{m_\pi})^2$    by using Eq.(\ref{phaseeq}) (red band)  compared with the phase shift  given by effective range expansion formula (black band), Eq.(\ref{effexpanrange}), where $a_0 m_\pi = 0.1019 \pm 0.0088$   and  $r_0 m_\pi = 9.0 \pm 2.4$ are taken from \cite{Horz:2019rrn}. For the reference, the physical phase shift by using parameterization given in \cite{Kaminski:2006qe} is also plotted (blue band). }\label{phaseplot}
\end{center}
\end{figure}

 The two-pion scattering parameters, the scattering length $a_0$ and effective range $r_0$,  can immediately be produced once $V_R(\mu)$ is determined from  the fit.  $a_0$ and $r_0$ are usually defined by  the effective range expansion at low energies:
\begin{equation}
p \cot \delta (E  ) = - \frac{1}{a_0} + \frac{r_0}{2} p^2 + \mathcal{O} (p^4),\label{effexpanrange}
\end{equation}
where 
$$
p = \frac{1}{2} \sqrt{E^2 - 4 m_\pi^2}
$$
is relative momentum of  the pions  in the CM frame.
 On the other hand, as detailed in Appendix \ref{2pionamp},    the isospin-$2$ $S$-wave phase shift of  $\pi^+ \pi^+$ scattering through a pair-wise contact potential is given by 
\begin{equation}
\cot \delta^{(2)}_0 (E  )  = - \frac{16 \pi }{ \sqrt{1 - \frac{4 m_\pi^2}{ E^2 } } }  \left [\frac{1}{V_R (\mu)}  - Re G ( E  )  + G (  \mu  )   \right ], \label{phaseeq}
\end{equation}
where  
$$
G (  E  ) =  \frac{\sqrt{1 - \frac{4 m_\pi^2}{ E^2 } }}{16 \pi^2} \ln \frac  {\sqrt{1 - \frac{4 m_\pi^2}{ E^2 } } + 1 }{\sqrt{1 - \frac{4 m_\pi^2}{ E^2 } } -1}  \, .
$$
 Comparing Eqs.~\eqref{effexpanrange} and \eqref{phaseeq}, one finds $a_0$ and $r_0$ in terms of $V_R(\mu)$:  
\begin{align}
\frac{1}{a_0 m_\pi} &=16 \pi  \left [ \frac{1}{ V_R (\mu) } +   G (  \mu  )  \right ],  \nonumber \\ 
r_0 m_\pi &= \frac{1}{a_0 m_\pi}  + \frac{4}{\pi} . \label{scattparameters}
\end{align}
The values of $\pi^+ \pi^+$ scattering length $a_0$ and  effective range $r_0$  
extracted in this study are compared in Table \ref{scatlength} with the values given by other works. Using Eq.(\ref{phaseeq}), the $\pi^+ \pi^+$ $S$-wave phase  shifts are plotted in  Fig.~\ref{phaseplot}.

Treating $V_R(\mu)$ as a free parameter,  one can produce the CM-frame $2\pi^+$ and $3\pi^+$ energy  spectra in  $A_1$ and $E$ irreps  with Eqs.~\eqref{2bqcirrep} and \eqref{3bqcirrep}, at lattice spacing $a=0.324 \mbox{GeV}^{-1}$ and pion mass $m_\pi =0.2 \mbox{GeV}$. The  spectra are  then matched to  the lattice results reported in Ref.~\cite{Horz:2019rrn},  as shown in Fig.~\ref{E2b3bplot}. The renormalization scale is chosen at $\mu = 0 \mbox{GeV}$, so that $G(\mu) = \frac{1}{8\pi^2}$  is real. The value of coupling constant of contact interaction is extracted,
\begin{equation}
V_R (0) \sim 17.5 \pm 12.5 .
\end{equation}
As illustrated in  Fig.~\ref{E2b3bplot}, with  only  one parameter,  
Eqs.~\eqref{2bqcirrep} and \eqref{2bqcirrep} struggle to match the lattice results,  so the fit yields a large  error on  the value of $V_R(0)$.  This   suggests that    more sophisticated pair-wise interactions and/or three-body interactions may be  needed.

 \begin{table}[htp]
\caption{ $\pi^+ \pi^+$  scattering length $a_0$ and  effective range $r_0$.}
\begin{center}
\begin{tabular}{|c |c|c|c|}
\hline
& $a_0 m_\pi$ & $r_0 m_\pi$ & $m_\pi (\mbox{GeV})$ \\
\hline
This work & $0.28 \pm 0.17$ &  $ 4.8 \pm 2.1$ & $0.200$ \\
\hline
Lattice in \cite{Horz:2019rrn} & $0.1019 \pm 0.0088$ &  $ 9.0 \pm 2.4$ & $0.200$ \\
\hline 
Lattice in   \cite{Dudek:2012gj}  & $ 0.307 \pm 0.013$ & $- 0.26 \pm 0.13$ & $0.396$  \\
\hline 
Analysis in   \cite{Blanton:2019vdk}  & $ 0.090 \pm 0.006$ & $28.78 \pm 0.89$ & $0.200$  \\
\hline 
\end{tabular}
\end{center}
\label{scatlength}
\end{table}

\section{Summary and outlook}\label{summary}
 Based on  the variational approach combined with the Faddeev method  proposed in Refs.~\cite{Guo:2018ibd,Guo:2019hih,Guo:2019ogp,Guo:2020wbl}, the relativistic $2\pi^+$ and $3\pi^+$ dynamics in finite volume  are  presented in    the paper.  The presentation of multi-$\pi^+$ dynamics  included both pair-wise and three-body interactions,   and the effects of finite lattice spacing and  projections onto irreps of the cubic lattice group were also discussed. The quantization conditions  were used to  analyze the lattice data published in \cite{Horz:2019rrn}.  In  the present work, only contact pair-wise interactions  were employed in our analysis  and, as a result,  the renormalized coupling strength  was the sole free parameter. The scattering length $a_0$ and  effective range $r_0$  were obtained,  however, with rather large uncertainty:   $a_0 m_\pi = 0.28 \pm 0.17$ and $r_0 m_\pi = 4.8 \pm 2.1$, which  might be  attributed to  the lack of accuracy in describing long-range interactions among the $\pi^+$'s by only a contact pair-wise interaction. 

 The idea of our framework is somewhat similar to that of Refs.~\cite{Hammer:2017uqm, Hammer:2017kms} in the sense both used an interaction to connect finite-volume and infinite-volume dynamics. But the implementations differ in many details. Relativistic kinematics of the pion and more lattice effects are considered here.

With the framework and computational facility developed here, we can improve readily the interactions by supplanting more sophisticated presentations for the two-body and three-body amplitudes. For instance, chiral perturbation theory can be employed systematically. With more degrees of freedom in parameterizing the interactions. It needs more efforts to incorporate a non-zero three-body force, but it can be done in the presented framework. These refinements on the interactions point to the future direction of this line of our research.

\bigskip

\begin{acknowledgements}
 We   acknowledge support from the Department of Physics and Engineering, California State University, Bakersfield, CA.   This research was supported in part by the National Science Foundation under Grant No. NSF PHY-1748958.   P.G.   acknowledges GPU computing resources (http://complab.cs.csubak.edu) from  the Department of Computer and Electrical Engineering and Computer Science at California State University-Bakersfield   made available for conducting the research reported in this work.  B.L. acknowledges support     by the National Science Foundation of China under Grant Nos.11775148 and 11735003.
 \end{acknowledgements}

\appendix

\section{Reduction of Bethe-Salpeter equation to relativistic $3\pi^+$ Lippmann-Schwinger equation }\label{reductionBS}
 
 Consider the general form of Bethe-Salpeter (BS) equation \cite{Salpeter:1951sz} for   three-scalar-particle bound states: 
\begin{align}
&  \psi_{BS} (p_1, p_2) = \frac{(-i)^2}{(p_1^2 - m_\pi^2)  (p_2^2 - m_\pi^2)  (p_3^2 - m_\pi^2) }  \nonumber \\
 & \times \int \frac{ d^4 p'_1}{(2\pi)^4}   \frac{ d^4 p'_2}{(2\pi)^4} I (p'_1 - p_1,  p'_2 - p_2) \psi_{BS} (p'_1 , p'_2 ) ,
\end{align}
where $p_i = (p_{i0}, \mathbf{ p}_i)$   are the four momenta of three particles,   the three-body BS wave function is labeled only by two independent particle momenta, $p_1$ and $p_2$, since the momenta of three particles are constrained by energy-momentum conservation, $p_3=P-p_1-p_2$.     Assuming  "instantaneous interaction  kernel", $I (p_1, p_2 )   = I(\mathbf{ p}_1, \mathbf{ p}_2 )$, and introducing Lippmann-Schwinger equation wave function, $\psi  (\mathbf{ p}_1, \mathbf{ p}_2)  = \int \frac{d p_{1 0}}{2\pi}  \frac{d p_{2 0}}{2\pi} \psi_{BS} (p_1, p_2) $, hence, we get
\begin{align}
&  \psi  (\mathbf{ p}_1, \mathbf{ p}_2)  =  \int \frac{d p_{1 0}}{2\pi}  \frac{d p_{2 0}}{2\pi}  \frac{(-i)^2}{(p_1^2 - m_\pi^2)  (p_2^2 - m_\pi^2)  (p_3^2 - m_\pi^2) }  \nonumber \\
 & \times \int \frac{ d  \mathbf{ p}'_1}{(2\pi)^3}   \frac{ d \mathbf{ p}'_2}{(2\pi)^3}  I(\mathbf{ p}'_1-\mathbf{ p}_1, \mathbf{ p}'_2-\mathbf{ p}_2 ) \psi  (\mathbf{ p}'_1, \mathbf{ p}'_2) .
\end{align}
The integration over propagators can be carried out,
\begin{align}
&  (2\pi)^6 G^{(\mathbf{ P})} ( \mathbf{ p}_1, \mathbf{ p}_2; E) \nonumber \\
& =  \int \frac{d p_{10}}{2\pi}  \frac{d p_{20}}{2\pi}  \frac{(-i)^2}{(p_1^2 - m_\pi^2)  (p_2^2 - m_\pi^2)  (p_3^2 - m_\pi^2) } \nonumber \\
&  =    \frac{1}{2E_{\mathbf{ p}_1} 2 E_{\mathbf{ p}_2} 2 E_{\mathbf{ p}_3}} \frac{2 (E_{\mathbf{ p}_1}+E_{\mathbf{ p}_2} +E_{\mathbf{ p}_3})}{E^2- (E_{\mathbf{ p}_1}+E_{\mathbf{ p}_2} +E_{\mathbf{ p}_3})^2}    ,
\end{align}
where  $E_{\mathbf{ p}_i} = \sqrt{  \mathbf{ p}_{i}^2+m_\pi^2}$.

The interactions of $3\pi^+$ consist of   pair-wise interactions and  three-body interaction potential, 
\begin{align}
& I(\mathbf{ p}'_1-\mathbf{ p}_1, \mathbf{ p}'_2-\mathbf{ p}_2 )  = \widetilde{V}_{(4)}(\mathbf{ p}'_1-\mathbf{ p}_1, \mathbf{ p}'_2-\mathbf{ p}_2 )  \nonumber \\
& +  \sum_{ k =1}^{ 3} 2 E_{\mathbf{ p}_k} (2\pi)^3 \delta(\mathbf{ p}'_k -\mathbf{ p}_k  ) \widetilde{ V} (  | \mathbf{ q}'_{ij} - \mathbf{ q}_{ij}  |  )|_{k\neq i \neq j}    ,
\end{align}
where symbols $\widetilde{V}$ and $\widetilde{V}_{(4)}$   are used to denote pair-wise interaction and three-body force respectively in consistent  with the conventions used in Section \ref{3bdynamics}. $ \mathbf{ q}_{ij}   = \frac{\mathbf{ p}_i - \mathbf{ p}_j}{2}$ refers to the relative momentum between i-th and j-th pions.
The relativistic kinematic factors $\langle \mathbf{ p}_k | \mathbf{ p}'_k \rangle=2 E_{\mathbf{ p}_k} (2\pi)^3 \delta(\mathbf{ p}'_k -\mathbf{ p}_k  )$ emerge   when k-th particle  is free propagating and  not involved in the  interaction.  The relativistic $3\pi^+$   Lippmann-Schwinger equation  is thus given explicitly by
 \begin{align}
 &  \psi  (\mathbf{ p}_1, \mathbf{ p}_2)   =   G^{(\mathbf{ P})} ( \mathbf{ p}_1, \mathbf{ p}_2; E)   \nonumber \\
  &  \times \bigg [      2 E_{\mathbf{ p}_1}  (2\pi)^3   \int  d  \mathbf{ p}'_2    \widetilde{V} (  | \mathbf{ p}'_{2} - \mathbf{ p}_{2} |  )   \psi  (\mathbf{ p}_1, \mathbf{ p}'_2)  \nonumber \\
  &     \quad +   2 E_{\mathbf{ p}_2} (2\pi)^3     \int   d  \mathbf{ p}'_1    \widetilde{V} (  | \mathbf{ p}'_{1} - \mathbf{ p}_{1}  |  )     \psi  (\mathbf{ p}'_1, \mathbf{ p}_2)  \nonumber \\
  &    \quad   +    2 E_{\mathbf{ p}_3}   (2\pi)^3   \int  d  \mathbf{ p}'_1        \widetilde{V} (  | \mathbf{ p}'_{1} - \mathbf{ p}_{1}  |  )     \psi  (\mathbf{ p}'_1,  \mathbf{ p}_1 + \mathbf{ p}_2 - \mathbf{ p}'_1) \nonumber \\
  & \quad +  \int  d  \mathbf{ p}'_1    d \mathbf{ p}'_2   \widetilde{V}_{(4)}(\mathbf{ p}'_1-\mathbf{ p}_1, \mathbf{ p}'_2-\mathbf{ p}_2 )   \psi  (\mathbf{ p}'_1, \mathbf{ p}'_2)   \bigg ] .
\end{align}

\section{$2\pi$ scattering amplitude in infinite volume with   contact interaction }\label{2pionamp}
With  the contact interaction, 
$$V (r) = V_0 \delta (\mathbf{ r}) \, ,$$
the relativistic Lippmann-Schwinger equation of $2\pi$ system  can be solved analytically, so the  scattering wave function of $2\pi$ in infinite volume and in CM frame is given by 
 \begin{equation}
 \phi_\mathbf{ p} (\mathbf{ r}) =e^{i \mathbf{ p} \cdot \mathbf{ r}} + G^{(\mathbf{ 0})}(\mathbf{ r} ; E  ) V_0  \phi_\mathbf{ p} (\mathbf{ 0}),  
 \end{equation}  
 where $\mathbf{ r}$ and $\mathbf{ p}$ are the relative coordinate and   momentum of  two pions  respectively.  The total energy of two pions is related to $\mathbf{ p}$ by  $E =2 E_\mathbf{ p} =2  \sqrt{\mathbf{ p}^2 + m_\pi^2 }$.
 The two pions Green's function in CM frame in infinite volume is given by
 \begin{equation}
 G^{(\mathbf{ 0})}  (\mathbf{ r}  ; E )  =  \int    \frac{d  \mathbf{ q} }{(2\pi)^3} \frac{1  }{E_\mathbf{ q}   } \frac{e^{i \mathbf{ q}    \cdot \mathbf{ r}}  }{E^2- (2 E_\mathbf{ q}  )^2 } .
 \end{equation}
 The two pions amplitude in CM frame is thus given by,
\begin{equation}
t  ( E ) = -   V_0   \phi_\mathbf{ p} (\mathbf{ 0}) = - \frac{1}{\frac{1}{V_0}  - G^{(\mathbf{ 0})}  (\mathbf{ 0}  ; E  ) } .
\end{equation}
 The Green's function $G^{(\mathbf{ 0})}  (\mathbf{ 0}  ; E )$ diverge at $\mathbf{ r}=\mathbf{ 0}$, a cutoff $\Lambda$ on momentum integration may be introduced to regularize ultraviolet divergence, as $\Lambda \rightarrow \infty$ hence we find
 \begin{equation}
  \int^\Lambda    \frac{d  \mathbf{ q} }{(2\pi)^3} \frac{1  }{E_\mathbf{ q}   } \frac{1  }{E^2- (2 E_\mathbf{ q}  )^2 } = G ( E )  -  \frac{1}{8\pi^2}  \ln \frac{2 \Lambda}{m_\pi}   ,
 \end{equation}
 where
  \begin{equation}
G ( E ) =   \frac{\sqrt{1 - \frac{4 m_\pi^2}{ E^2 } }}{16 \pi^2} \ln \frac  {\sqrt{1 - \frac{4 m_\pi^2}{ E^2 } } + 1 }{\sqrt{1 - \frac{4 m_\pi^2}{ E^2 } } -1}   .
 \end{equation}
The imaginary part of  function $G ( E  ) $ is   non-zero only above threshold of two pions: $Im G ( E  )  = -   \frac{1}{16 \pi} \sqrt{1 - \frac{4 m_\pi^2}{ E^2 } }$ for $E > 2 m_\pi$, and zero otherwise. The ultraviolet divergence may be absorbed by redefining bare coupling, $V_0$, 
\begin{equation}
\frac{1}{V_0} = \frac{1}{V_R (\mu)} + G (  \mu  ) -  \frac{1}{8\pi^2}  \ln \frac{2 \Lambda}{m_\pi} ,
\end{equation} 
 where $V_R (\mu)$ stands for the renormalized coupling strength at a renormalization scale $\mu$. $\mu$ will be chosen below two pions threshold, so that $G (  \mu  ) $ is real. Therefore, the cutoff dependence is cancelled out completely, and  renormalized scattering amplitude is now given by
\begin{equation}
t  ( E ) = - \frac{1}{ \frac{1}{V_R (\mu)}   -  G ( E  )  + G (  \mu  )  } , \ \ \mu < 2 m_\pi. 
\end{equation}
The phase shift is defined by $t  ( E )  = \frac{16 \pi }{ \sqrt{1 - \frac{4 m_\pi^2}{ E^2 } } }  \frac{1}{\cot \delta (E  ) - i }$, hence we obtain 
\begin{equation}
\cot \delta (E  )  = - \frac{16 \pi }{ \sqrt{1 - \frac{4 m_\pi^2}{ E^2 } } }  \left [\frac{1}{V_R (\mu)}  - Re G ( E  )  + G (  \mu  )   \right ].
\end{equation}
Expanding phase shift near threshold $p  = | \mathbf{ p}| = \frac{1}{2} \sqrt{ E^2 - 4 m_\pi^2} \sim 0$, we find
\begin{equation}
p \cot \delta (E  ) = - \frac{1}{a_0} + \frac{r_0}{2} p^2 + \mathcal{O} (p^4),
\end{equation}
where the effective expansion parameters, $a_0$ and $r_0$ are given by
\begin{align}
\frac{1}{a_0 m_\pi} &=16 \pi  \left [ \frac{1}{ V_R (\mu) } +   G (  \mu  )  \right ], \nonumber \\
r_0 m_\pi & = 16 \pi  \left [ \frac{1}{ V_R (\mu) } +   G (  \mu  )  \right ] + \frac{4}{\pi}.
\end{align}

\section{Character Table for the octahedral group $O_h$  }\label{chartable}

 The octahedral group $O_h$ is  the direct product of  proper  rotational group $O$  that  rotates a cube into itself and spatial inversion $I$: 
 $$
 O_h = O \times I \, .
 $$
   $O_h$ contains 48 elements, including 24  elements of rotational group $O$  and 24 elements of combined operation of inversion and rotations: $I g$ where $g \in O$.  The   matrix  representations of 24 proper  rotational group $O$ are given by
\begin{widetext}
\begin{align}
&  E = \begin{bmatrix} 1& 0 &0 \\ 0 & 1 & 0 \\ 0& 0&1 \end{bmatrix} , \ \ \ \ C_{2x} = \begin{bmatrix} 1& 0 &0 \\ 0 & -1 & 0 \\ 0& 0& -1 \end{bmatrix} , \ \ \ \  C_{2y} = \begin{bmatrix} -1& 0 &0 \\ 0 & 1 & 0 \\ 0& 0& -1 \end{bmatrix} , \ \ \ \  C_{2z} = \begin{bmatrix} -1& 0 &0 \\ 0 & -1 & 0 \\ 0& 0& 1 \end{bmatrix} , \nonumber \\
& C_{4x} = \begin{bmatrix} 1& 0 &0 \\ 0 & 0 & 1 \\ 0& -1&0  \end{bmatrix} , \ \ \ \  C_{4y} = \begin{bmatrix} 0& 0 &-1 \\ 0 & 1 & 0 \\ 1& 0& 0 \end{bmatrix} , \ \ \ \   C_{4z} = \begin{bmatrix} 0& 1 &0 \\ -1 & 0 & 0 \\ 0& 0& 1 \end{bmatrix} , \ \ \ \   C^{-1}_{4x} =  \begin{bmatrix} 1& 0 &0 \\ 0 & 0 & -1 \\ 0& 1&0  \end{bmatrix} , \nonumber \\
&   C^{-1}_{4y} = \begin{bmatrix} 0& 0 &1 \\ 0 & 1 & 0 \\ -1& 0& 0 \end{bmatrix} , \ \ \ \   C^{-1}_{4z} = \begin{bmatrix} 0& -1 &0 \\ 1 & 0 & 0 \\ 0& 0& 1 \end{bmatrix},  \ \ \ \   C_{2a} = \begin{bmatrix} 0& 1 &0 \\ 1 & 0 & 0 \\ 0& 0& -1 \end{bmatrix} , \ \ \ \   C_{2b} = \begin{bmatrix} 0& -1 &0 \\ -1 & 0 & 0 \\ 0& 0& -1 \end{bmatrix}, \nonumber \\ %\\
%\end{align}
%\begin{align}
&  C_{2c} = \begin{bmatrix} 0& 0 &1 \\ 0 &- 1 & 0 \\ 1& 0& 0 \end{bmatrix} , \ \ \ \ C_{2d} = \begin{bmatrix} 0&  0&-1 \\ 0 & -1 & 0 \\ -1& 0& 0 \end{bmatrix},  \ \ \ \   C_{2e} = \begin{bmatrix} -1& 0 &0 \\ 0 & 0 & 1 \\ 0& 1& 0 \end{bmatrix} , \ \ \ \   C_{2f} = \begin{bmatrix} -1& 0 &0 \\ 0 & 0 & -1 \\ 0& -1& 0 \end{bmatrix}, \nonumber \\
&   C_{3\alpha} = \begin{bmatrix} 0& 1 &0 \\ 0 &0 & -1 \\ -1& 0& 0 \end{bmatrix} , \ \ \ \   C_{3\beta} = \begin{bmatrix} 0&  -1& 0 \\ 0 & 0 & -1 \\ 1& 0& 0 \end{bmatrix},  \ \ \ \   C_{3\gamma} = \begin{bmatrix}  0 & -1 &0 \\ 0 & 0 & 1 \\ -1& 0& 0 \end{bmatrix} , \ \ \ \   C_{3\delta} = \begin{bmatrix} 0& 1 &0 \\ 0 & 0 & 1 \\ 1& 0 & 0 \end{bmatrix}, \nonumber \\
&   C^{-1}_{3\alpha} = \begin{bmatrix} 0& 0 &-1 \\ 1 &0 & 0 \\ 0& -1& 0 \end{bmatrix} , \ \ \ \   C^{-1}_{3\beta} = \begin{bmatrix} 0&  0& 1 \\ -1 & 0 & 0 \\ 0& -1& 0 \end{bmatrix},  \ \ \ \   C^{-1}_{3\gamma} = \begin{bmatrix}  0 & 0 &-1 \\ -1 & 0 & 0 \\ 0& 1& 0 \end{bmatrix} , \ \ \ \   C^{-1}_{3\delta} = \begin{bmatrix} 0& 0 &1 \\ 1 & 0 & 0 \\ 0& 1 & 0 \end{bmatrix},
\end{align}
\end{widetext}
and  the inversion matrix,
\begin{align}
&  I = \begin{bmatrix} -1& 0 &0 \\ 0 & -1 & 0 \\ 0& 0&- 1 \end{bmatrix} .
\end{align}

 \begin{table}[htp]
\caption{Character table for  irreps of   the octahedral group  $O_h$ with positive parity }
\begin{center}
\begin{tabular}{|c|c|c|c|c|c|}
\hline
  & $ \chi (\mathcal{C}_1, \mathcal{C}_6)  $ & $\chi (\mathcal{C}_2, \mathcal{C}_7) $ & $\chi(\mathcal{C}_3,\mathcal{C}_8) $   & $\chi (\mathcal{C}_4,\mathcal{C}_9)  $   & $\chi (\mathcal{C}_5,\mathcal{C}_{10})  $  \\
\hline
$A^+_1$ & 1&1&1&1&1  \\
\hline
$A^+_2$ & 1&1&1&-1&-1  \\
\hline
$E^+$ & 2&-1&2&0&0  \\
\hline
$T^+_1$ & 3&0&-1&1&-1  \\
\hline
$T^+_2$ & 3&0&-1&-1&1  \\
\hline
\end{tabular}
\end{center}
\label{fullcharOh}
\end{table}

All 48 elements of the octahedral group $O_h$  are usually grouped into different conjugacy classes,  and the members within the same conjugacy class share the same   character value  for a  given irrep. Using the same convention as used in  \cite{Cornwell:1997ke}, ten classes of the octahedral group $O_h$ are named as $\mathcal{C}_i$ where $i=1,\cdots, 10$, they are associated with 48 elements by
\begin{align}
& \mathcal{C}_1=E, \ \  \ \  \mathcal{C}_2= \left ( \begin{aligned} & C_{3\alpha}, C_{3\beta},C_{3\gamma},C_{3\delta}, \\ 
 &C^{-1}_{3\alpha}, C^{-1}_{3\beta},C^{-1}_{3\gamma},C^{-1}_{3\delta} \end{aligned} \right ) , \nonumber \\
&  \mathcal{C}_3= (C_{2x}, C_{2y}, C_{2z}), \ \ \ \  \mathcal{C}_4= \left ( \begin{aligned} & C_{4x}, C_{4y}, C_{4z} , \\
 & C^{-1}_{4x}, C^{-1}_{4y}, C^{-1}_{4z} \end{aligned} \right ) , \nonumber \\
 & \mathcal{C}_5=  \left (  C_{2a}, C_{2b}, C_{2c},   C_{2d}, C_{2e},C_{2f}  \right ), 
\end{align} 
 and $  \mathcal{C}_i = I  \mathcal{C}_{(i-5)}$ for $i=6,\cdots , 10$. The character table for all irreps of  the octahedral group $O_h$ with positive parity quantum number  is given in Table \ref{fullcharOh}.

\section{Non-relativistic three-particle dynamics in finite volume }\label{nonrel3piondynamics}
The dynamics of three non-relativistic identical bosonic particles in finite volume is described by the 
\begin{align}
& \left [  2 m E+ \sum_{i=1}^3  \nabla_i^2 -   \sum_{k=1}^3   U ( r_{ij})  -  U_{(4)}  (\mathbf{ r}_{13},\mathbf{ r}_{23})  \right ]    \nonumber \\
&       \times \Phi(\mathbf{ x}_1,\mathbf{ x}_2,\mathbf{ x}_3)  =0,  \ \ \ \ i \neq j \neq k, \label{3bschrodinger}
\end{align}
where   $m $ is the mass of identical bosons. $\mathbf{ x}_i$ denotes the position of i-th  particle, and $\mathbf{ r}_{ij} = \mathbf{ x}_i - \mathbf{ x}_j$ is relative coordinate between i-th and j-th particles. The pair-wise interaction between i-th and j-th particles is described by $U  ( r_{ij})    $, and $U_{(4)}  (\mathbf{ r}_{13},\mathbf{ r}_{23})   $ represents the three-body interaction among all particles.   Both pair-wise and three-body interactions in finite volume are assumed to be short-range and  periodic, that is to say
\begin{align}
& U  ( r ) = U  (  |\mathbf{ r} + \mathbf{ n} L| ), \ \ \mathbf{ n} \in \mathbb{Z}^3, \nonumber \\
&U_{(4)}  (\mathbf{ r}_{13},\mathbf{ r}_{23})  = U_{(4)}  (\mathbf{ r}_{13} + \mathbf{ n}_1 L ,\mathbf{ r}_{23} +  \mathbf{ n}_2 L) ,   \ \  \mathbf{ n}_{1,2} \in \mathbb{Z}^3,
\end{align}
where $L$ is the size of the cubic lattice. Therefore finite volume three-particle wave function must also satisfy periodic boundary condition,
\begin{equation}
  \Phi(\mathbf{ x}_1,\mathbf{ x}_2,\mathbf{ x}_3)  = \Phi(\mathbf{ x}_1+ \mathbf{ n}_{\mathbf{ x}_1},\mathbf{ x}_2 + \mathbf{ n}_{\mathbf{ x}_2} ,\mathbf{ x}_3 + \mathbf{ n}_{\mathbf{ x}_3})   , 
\end{equation}
where $ \mathbf{ n}_{\mathbf{ x}_i} \in \mathbb{Z}^3$. As suggested in \cite{Guo:2019hih,Guo:2019ogp}, it may be more convenient to consider the integral representation of Eq.(\ref{3bschrodinger}),
\begin{align}
&  \Phi(\mathbf{ x}_1,\mathbf{ x}_2,\mathbf{ x}_3)  = \int_{L^3} \prod_{i=1}^3 d \mathbf{ x}'_i   \frac{1}{L^9} \sum_{ \mathbf{ p}_1, \mathbf{ p}_2, \mathbf{ p}_3 }  \frac{e^{ i \sum_{i=1}^3 \mathbf{ p}_i \cdot  ( \mathbf{ x}_i -  \mathbf{ x}'_i ) }}{ 2 m E - \sum_{i=1}^3  \mathbf{ p}^2_i } \nonumber \\
& \times    \left [   \sum_{k=1}^3 U  ( r'_{ij})  +  U_{(4)}  (\mathbf{ r}'_{13},\mathbf{ r}'_{23})  \right ]  \Phi(\mathbf{ x}'_1,\mathbf{ x}'_2,\mathbf{ x}'_3)  ,  \label{3bnonrelLSeqfull}
\end{align}
where $\mathbf{ p}_{1,2,3} \in \frac{2\pi \mathbf{ n}}{L}$, $\mathbf{ n} \in \mathbb{Z}^3$.
The center of mass motion of three-particle system can be factorized by 
\begin{equation}
 \Phi(\mathbf{ x}_1,\mathbf{ x}_2,\mathbf{ x}_3)  = e^{ i \mathbf{ P} \cdot \mathbf{ R}}  \phi(\mathbf{ r}_{13},\mathbf{ r}_{23}) ,
\end{equation}
where $\mathbf{ R} = \frac{\mathbf{ x}_1+ \mathbf{ x}_2+\mathbf{ x}_3}{3} = \frac{\mathbf{ r}_{13} + \mathbf{ r}_{23}}{3} +\mathbf{ x}_3$ is center of mass position of three-particle system, and $\mathbf{ P} = \frac{2\pi}{L} \mathbf{ d}$ with $\mathbf{ d} \in \mathbb{Z}^3$ stands for the total momentum of three-particle in a periodic cubic box.  $ \phi(\mathbf{ r}_{13},\mathbf{ r}_{23}) $ is the wave function that is associated with the internal motion of three particles, and it satisfies periodic boundary condition,
\begin{equation}
 \phi(\mathbf{ r}_{13} + \mathbf{ n}_1 L,\mathbf{ r}_{23} +\mathbf{  n}_2 L)  = e^{  - i \frac{\mathbf{ P}}{3} \cdot ( \mathbf{ n}_1 L + \mathbf{ n}_2 L) } \phi(\mathbf{ r}_{13},\mathbf{ r}_{23}) ,
\end{equation}
where $ \mathbf{ n}_{1,2} \in \mathbb{Z}^3$.     We remark that  in this work we use $(\mathbf{ r}_{13},\mathbf{ r}_{23})$  to describe the internal motion of three particles, and $\mathbf{ x}_3$ is thus associated to CM motion.  So that $ \int_{L^3} \prod_{i=1}^3 d \mathbf{ x}'_i   = \int_{L^3}  d \mathbf{ r}'_{13}  d \mathbf{ r}'_{23}   d \mathbf{ x}'_3    $, and it resembles a two-light and one heavy three-body atomic system.   Integrating out CM motion,
\begin{equation}
\int_{L^3}  d \mathbf{ x}'_3  e^{i (\mathbf{ P} - \mathbf{ p}_1- \mathbf{ p}_2 - \mathbf{ p}_3)\cdot \mathbf{ x}'_3} =  L^3 \delta_{\mathbf{ p}_3, \mathbf{ P} - \mathbf{ p}_1- \mathbf{ p}_2},
\end{equation}
 the three-particle Lippmann-Schwinger equation, Eq.(\ref{3bnonrelLSeqfull}), now is reduced to
\begin{align}
& \phi(\mathbf{ r}_{13},\mathbf{ r}_{23})  = \int_{L^3}  d \mathbf{ r}'_{13}  d \mathbf{ r}'_{23}  G^{(\mathbf{ P})} ( \mathbf{ r}_{13}- \mathbf{ r}'_{13} ,\mathbf{ r}_{23} -\mathbf{ r}'_{23} ;E)  \nonumber \\
& \times   \left [   \sum_{k=1}^3 U  ( r'_{ij})  +  U_{(4)}  (\mathbf{ r}'_{13},\mathbf{ r}'_{23})  \right ]   \phi(\mathbf{ r}'_{13},\mathbf{ r}'_{23})  .  \label{3bnonrelLSeq}
\end{align}
The three-particle Green's function is defined by
\begin{align}
& G^{(\mathbf{ P})} ( \mathbf{ r}_{13} ,\mathbf{ r}_{23}  ;E)  \nonumber \\
&=    \sum_{ \mathbf{ p}_1, \mathbf{ p}_2  }   e^{ i  ( \mathbf{ p}_1- \frac{\mathbf{ P}}{3}) \cdot    \mathbf{ r}_{13}   } e^{ i  ( \mathbf{ p}_2 - \frac{\mathbf{ P}}{3}) \cdot    \mathbf{ r}_{23}   }  \widetilde{G}^{(\mathbf{ P})} ( \mathbf{ p}_{1} ,\mathbf{ p}_{2}  ;E)   , \nonumber \\
&\widetilde{G}^{(\mathbf{ P})} ( \mathbf{ p}_{1} ,\mathbf{ p}_{2}  ;E) =\frac{1}{L^6}   \frac{1}{ 2 m E - \sum_{i=1}^3  \mathbf{ p}^2_i } ,
\end{align}
where $\mathbf{ p}_{1,2} \in \frac{2\pi \mathbf{ n}}{L}$, $\mathbf{ n} \in \mathbb{Z}^3$, and $\mathbf{ p}_3 = \mathbf{ P}-\mathbf{ p}_{1}-\mathbf{ p}_{2}$.

Following  the same   procedures  as   described in Section \ref{3bdynamics}, only two independent scattering amplitudes are required due to exchange symmetry of three-particle wave function,
 \begin{align}
T_{(2)}^{(\mathbf{ P})}(\mathbf{ k}_1, \mathbf{ k}_2) = - & \int_{L^3} d \mathbf{ r}_{13}  d \mathbf{ r}_{23}  e^{-i (\mathbf{ k}_1 - \frac{\mathbf{ P}}{3}) \cdot \mathbf{ r}_{13} } e^{-i (\mathbf{ k}_2 - \frac{\mathbf{ P}}{3}) \cdot \mathbf{ r}_{23} }  \nonumber \\
 & \times  U (r_{13} )  \phi  (\mathbf{ r}_{13}, \mathbf{ r}_{23}) ,  \nonumber \\
  T_{(4)}^{(\mathbf{ P})}(\mathbf{ k}_1, \mathbf{ k}_2) = - & \int_{L^3} d \mathbf{ r}_{13}  d \mathbf{ r}_{23}  e^{-i (\mathbf{ k}_1 - \frac{\mathbf{ P}}{3}) \cdot \mathbf{ r}_{13} } e^{-i (\mathbf{ k}_2 - \frac{\mathbf{ P}}{3}) \cdot \mathbf{ r}_{23} }  \nonumber \\
 & \times  U_{(4)} (\mathbf{ r}_{13} ,\mathbf{ r}_{23} )  \phi  (\mathbf{ r}_{13}, \mathbf{ r}_{23}) . 
 \end{align}
Two amplitudes, $T_{(2)}^{(\mathbf{ P})}$ and $T_{(4)}^{(\mathbf{ P})}$, satisfy equations,
 \begin{align}
& T_{(2)}^{(\mathbf{ P})}(\mathbf{ k}_1, \mathbf{ k}_2)    =        \sum_{ 
              \mathbf{ p}_{1}    }      \widetilde{U}   ( | \mathbf{ k}_1- \mathbf{ p}_1   |) L^3  \widetilde{G}^{(\mathbf{ P})}  ( \mathbf{ p}_1 ,  \mathbf{ k}_2 ;E)   \nonumber \\
 &\times     \bigg [     T_{(2)}^{(\mathbf{ P})}(\mathbf{ p}_1, \mathbf{ k}_2) +   T_{(2)}^{(\mathbf{ P})}( \mathbf{ k}_2,\mathbf{ p}_1)   \nonumber \\
 & \quad  +    T_{(2)}^{(\mathbf{ P})}(\mathbf{ p}_1, \mathbf{ P}-\mathbf{ p}_1- \mathbf{ k}_2)   +   T_{(4)}^{(\mathbf{ P})}(\mathbf{ p}_1, \mathbf{ k}_2) \bigg ]  ,   \label{nonrelt2eq}
 \end{align}  
 and
   \begin{align}
 & T_{(4)}^{(\mathbf{ P})}(\mathbf{ k}_1, \mathbf{ k}_2)  
 =        \sum_{ 
              \mathbf{ p}_{1}   , \mathbf{ p}_{2}    }     \widetilde{U}_{(4)} ( \mathbf{ k}_1 - \mathbf{ p}_1,\mathbf{ k}_2 -  \mathbf{ p}_2 )    \nonumber \\
 &   \times     \widetilde{G}^{(\mathbf{ P})}  ( \mathbf{ p}_1 ,  \mathbf{ p}_2 ;E)    \bigg [     T_{(2)}^{(\mathbf{ P})}(\mathbf{ p}_1, \mathbf{ p}_2) +    T_{(2)}^{(\mathbf{ P})}(\mathbf{ p}_2, \mathbf{ p}_1)   \nonumber \\
 & \quad \quad\quad\quad \quad\quad \quad\quad     +   T_{(2)}^{(\mathbf{ P})}(\mathbf{ p}_1, \mathbf{ p}_3)    +   T_{(4)}^{(\mathbf{ P})}(\mathbf{ p}_1, \mathbf{ p}_2) \bigg ]  ,  \label{nonrelt4eq}
 \end{align}
  where $\mathbf{ p}_3 =  \mathbf{ P} - \mathbf{ p}_1- \mathbf{ p}_2$, and  $  \widetilde{U} $ and  $ \widetilde{U}_{(4)}  $ are the Fourier transform of interaction potentials  $U$ and $U_{(4)}$ respectively.  Non-relativistic three-particle dynamical equations, Eq.(\ref{nonrelt2eq}) and Eq.(\ref{nonrelt4eq}), resemble their relativistic counter parts, Eq.(\ref{3bLSred1}) and Eq.(\ref{3bLSred2}), excepts some relativistic kinematic factors.

\bibliography{ALL-REF.bib}

\end{document}